\documentclass[preprint2]{aastex}

\shorttitle{Evolution of Primordial Stars}
\shortauthors{Hirano, Umeda and Yoshida}

\begin{document}


\title{Evolution of Primordial Stars Powered by Dark Matter \\Annihilation up to the Main-Sequence Stage}


\author{Shingo Hirano}
\affil{Department of Astronomy, School of Science, University of Tokyo, Hongo Tokyo 113-0033, Japan}
\email{hirano@astron.s.u-tokyo.ac.jp}

\author{Hideyuki Umeda}
\affil{Department of Astronomy, School of Science, University of Tokyo, Hongo Tokyo 113-0033, Japan}
\email{umeda@astron.s.u-tokyo.ac.jp}

\and

\author{Naoki Yoshida}
\affil{IPMU, University of Tokyo, 5-1-5 Kashiwanoha, Kashiwa, Chiba 277-8583, Japan}
\email{naoki.yoshida@ipmu.jp}




\begin{abstract}
Primordial stars formed in the early universe are
thought to be hosted by compact dark matter (DM) halos. 
If DM consists of Weakly Interacting Massive Particles (WIMPs),
such stars may be powered by DM annihilation during 
the early phases of their evolutions.
We study the pre-main sequence evolutions of the primordial star
using a detailed stellar evolution code
under the assumption that the annihilation of adiabatically contracted 
WIMPs DM within the star provides a sufficient energy to sustain the 
stellar equilibrium. 
We follow the evolution of accreting stars
using several gas mass accretion rates derived from cosmological simulations.
We show that the stellar mass becomes very large, up to $900 - 1000 \ \mathrm{M_{\odot}}$
when the star reaches the main-sequence phase for a reasonable set of model 
parameters such as DM particle mass and the annihilation cross section.
During the dark star phase, the star expands over a thousand solar-radii,
while the surface temperature remains below $10^4 \ \mathrm{K}$. 
The energy generated by nuclear reactions is not dominant during this phase. 
We also study models with different gas mass accretion rates 
and the DM particle masses. 
All our models for different DM particle masses pass the dark star phase. 
The final mass of the dark stars 
is essentially unchanged for DM mass of $m_{\chi} \le 10 \ \mathrm{GeV}$.
Gravitational collapse of the massive dark stars will leave massive
black holes with mass as large as $1000 \ \mathrm{M_{\odot}}$ in the early universe. 
\end{abstract}





\section{Introduction}

The first stars in the universe may have contributed early cosmic
reionization and may have also enriched the inter-galactic medium with 
heavy elements such as carbon, oxygen and iron 
(see, e.g. \citet{bromm09} for a review). 
Future observations of the distant universe will exploit
large ground-based and space-borne telescopes such as
{\sl James Webb Space Telescope} and {\sl Thirty Meter Telescope}
to answer the important questions of how and when
the first stars were formed and how they affected the subsequent
evolution of the universe.

Recent theoretical studies based on cosmological simulations
suggest that the first stars are formed in dark matter (DM) halos with 
mass $10^5-10^6 \ \mathrm{M_{\odot}}$ 
at $z \sim 20-30$ \citep{tegmark97, yoshida03}.
In such a cosmological ``minihalo'', the gas cools and condenses 
by molecular hydrogen cooling 
to form a star-forming gas cloud.
The gravitationally unstable cloud further contracts,
and finally a proto-star is born inside it \citep{omukai98, yoshida08}. 
A unique characteristic of primordial star formation in the
standard cold dark matter model is that the star and its parent gas cloud 
are embedded at the center of the host dark halo. 
Thus the formation and evolution of primordial stars 
may be much affected by DM dynamically.

The nature of DM remains still unknown, but the best candidates
are thought to be Weakly Interacting Massive Particles, WIMPs.
WIMP DM such as neutralinos must have a large self-annihilation
cross-section in order to be a dominant component of DM
in the present-day universe. 
DM annihilation produces an enormous amount of energy, essentially
equal to the rest mass energy of the annihilated DM particles.
If DM density can be very large in a primordial 
gas cloud and in a primordial star,
the annihilation energy would act as an efficient heat input.
DM annihilation could even supply sufficient energy to 
support self-gravitation of a star. 
This is indeed a recently proposed new stellar phase, 
in which stars are powered 
by DM annihilation energy instead of that of the nuclear fusion 
\citep{spolyar08}. 
Hereafter we call such a star ``{\sl dark star}''.

\citet{spolyar08} suggest that, in a collapsing
primordial gas cloud, DM annihilation heating can win over
the radiative cooling, effectively halts further collapse of the cloud.
\citet{natarajan09} study the evolution of the DM density profile
during the first star formation
using cosmological simulations. They show that the DM density profile
indeed becomes very steep, as much as expected by adiabatic contraction models.

There have been many previous studies on dark stars. While most of them
assume constant stellar mass models \citep{iocco08, taoso08, yoon08}.
\citet{spolyar09} and \citet{umeda09} study 
dark star evolution with gas mass accretion.
By using an approximated stellar structure model,
\citet{spolyar09} follow dark star evolutions with
adiabatically contracted DM (ACDM). They show that the final stellar
mass becomes as large as $1000 \ \mathrm{M_{\odot}}$. 
\citet{umeda09} perform detailed, dynamically self-consistent 
stellar evolution calculations up to 
gravitational core-collapse. 
They show that the final mass can be as large as $10^{4-5} \ \mathrm{M_{\odot}}$ 
or more, if the star captures DM efficiently. However, the DM capture rate itself
is uncertain because it depends
on the unknown DM-baryon scattering cross section and also on the ambient
DM density during the dark star evolution. Thus the proposed formation
of inter-mediate massive black holes 
appears to occur only in very
particular cases.

An attractive feature of the dark star models is that
the possibility for the birth of massive stars, 
and black holes (BHs) as the end of such stars, in the early universe.
Super-massive BHs (SMBHs) existed already at $z \sim 6$, when the age
of the universe is less than one billion years old, 
but the formation mechanism of such early SMBHs remains unknown.
Gravitational collapse of a very massive dark star 
might provide a solution to this problem. 
We explore a possibility that the first stars
formed in DM halos can become very massive by being powered
by DM annihilation. To this end,
it is important to determine the final state of the dark star phase exactly, 
and to determine the final stellar mass.

In this paper, we study the evolution of an accreting dark star
using a detailed stellar evolution code as in \citet{umeda09}
but with the ACDM annihilation. We calculate the
density profile of ACDM using an analytical method
and determine the DM annihilation rate.
We then follow the evolution of the
dark star which grows in mass by gas accretion. The dark star model has
two parameters effectively; the gas mass accretion rate and
DM particle mass.
We calculate several models with different set of parameters and
investigate the effects on the final stellar mass.
Especially, we clarify the effect of the different mass accretion rates,
which were not explicitly shown in a similar work by \citet{spolyar09}.

The rest of the paper is organized as follows.
In section 2, we introduce our numerical calculation and explain how 
we implement DM annihilation in the stellar evolution.
Sec. 3 shows the results of our dark star models 
and we discuss the implications in Sec. 4. 
We give our concluding remarks in Sec. 5.

\section{Method}

We use a stellar evolution code with gas mass accretion 
developed by \citet{ohkubo06, ohkubo09, umeda09}. 
We implement energy generation by DM annihilation rate in the code
in the following manner.
We consider a primordial (proto-)star embedded at the center
of a small mass DM halos.
The DM density profile, ${\rho}_{\chi}$,
is calculated using the analytical method of \citet{blumenthal86}; 
when the primordial gas cools and collapses, DM particles  
are also ``dragged'' into the gravitational potential well. 
The adiabatic contraction model uses an adiabatic invariant. 
For spherically distributed DM and baryons
with a total mass of $M(R) = M_{\mathrm{gas}}(R) + M_{\mathrm{DM}}(R)$, 
an adiabatic invariant is
\begin{equation}
	M(R)R = const.
	\label{eq:blumenthal}
\end{equation}
\citet{freese09} compared the difference of DM density profile calculated 
using this method and a more accurate method developed by \citet{young80}. 
They conclude that the difference is not more than a factor of two. 

\begin{figure}
\begin{center}
\includegraphics{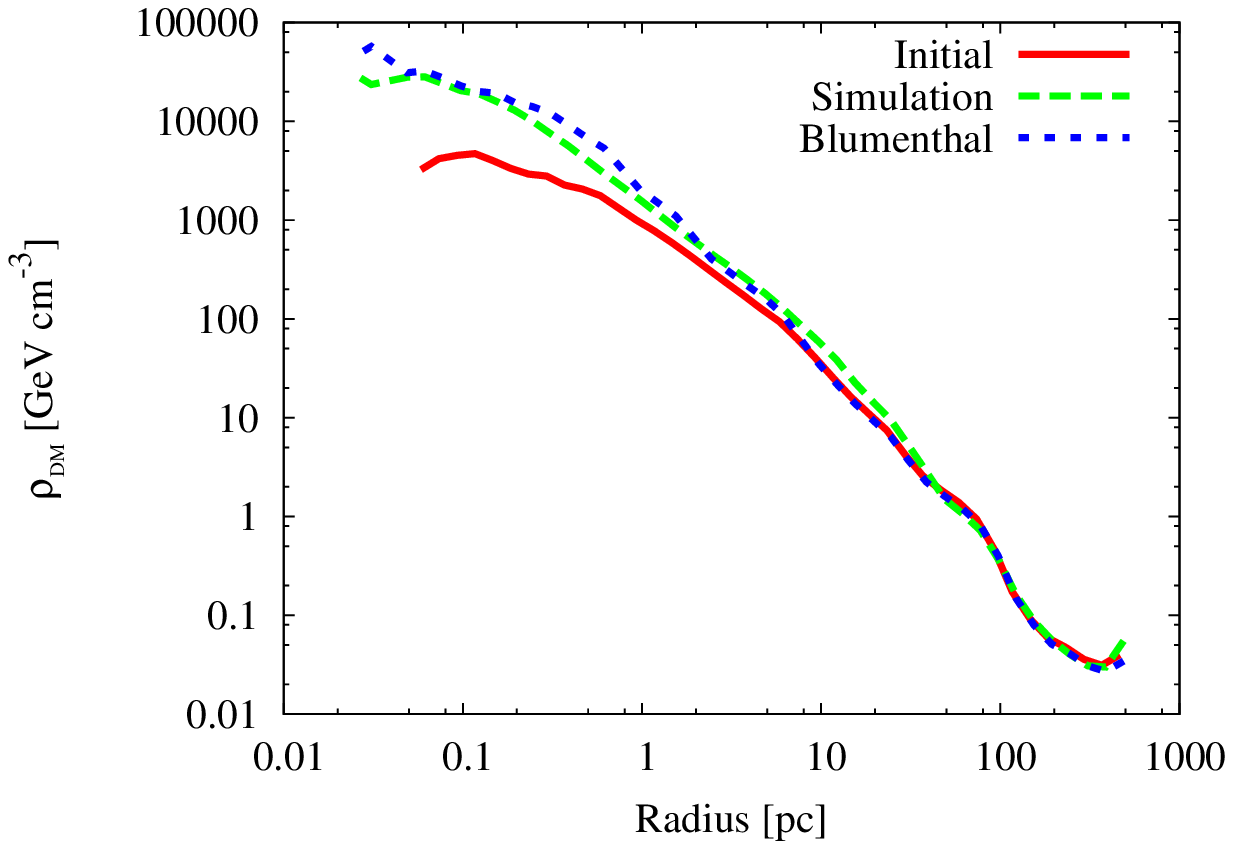}
\caption{
Dark matter density profile calculated by
an adiabatic contraction model.
The solid and dashed lines are calculated directly 
from our cosmological $N$-body simulation, 
whereas the dotted line is calculated from the initial profile data using the analytical method of \citet{blumenthal86}.
The dashed and dotted profiles agree remarkably well.}
\label{f1}
\end{center}
\end{figure}

We have performed direct numerical simulations of early structure formation 
to check the accuracy of this analytical method.
To this end we have used the parallel $N$-body/Smoothed Particle 
Hydrodynamics (SPH) solver 
GADGET-2 \citep{springel05} in its version suitably adopted to 
follow radiative cooling processes at very high densities \citep{yoshida06}. 
We have found that the DM density calculated by Blumenthal's method 
is indeed in good agreement with the result of our direct numerical
simulations. Figure \ref{f1} compares the results from the analytic model 
and the simulations.
Details will be presented elsewhere (Hirano et al., in preparation).
We note that, although these two results agree very well, confirming
the validity of the analytic model,
the simulations do not resolve the DM density down to 
length scales of astronomical units. 
Because the gas density profile itself evolves in a self-similar manner 
\citep{omukai98},
we assume that the DM continue to contract adiabatically and 
the evolution of DM density can be calculated by Blumenthal's method.

In this work, we only consider the DM density evolution
based on adiabatic contraction. In principle, DM annihilation can occur
even more efficiently in the star by capture. If the
density of the stellar core increases enough to scatter and trap DM particles,
the DM particles can rapidly sink toward the center, to self-annihilate
efficiently. During the dark star phase, however,
the effect of DM capture is unimportant because the star has expanded
extremely and the gas density is low.

\begin{figure}
\begin{center}
\includegraphics{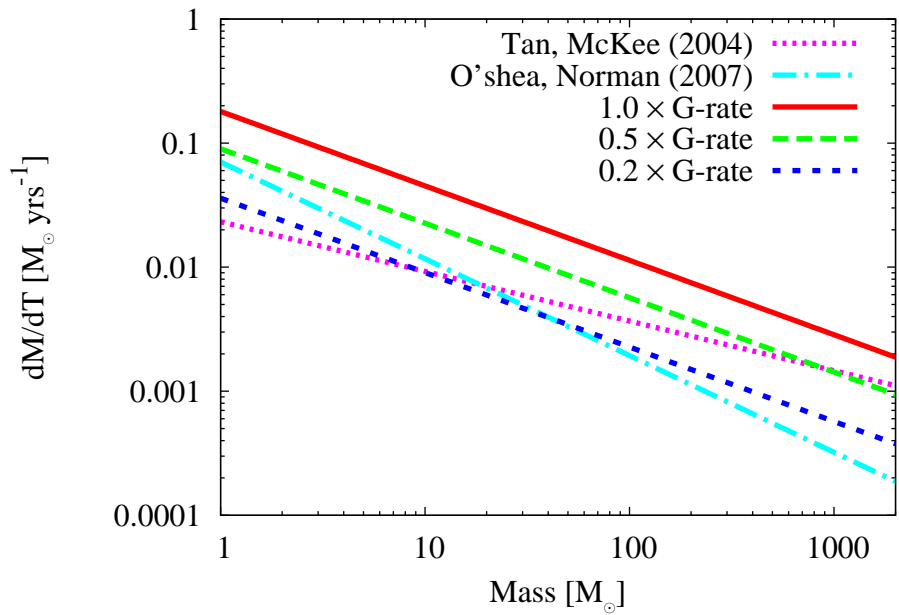}
\caption{
Gas mass accretion rates adopted in our stellar evolution calculations.
``G-rate'' is the accretion rate of Eq. \ref{eq:G-rate}.
We adopt three rates $1.0,\ 0.5$ and $0.2 \times $ G-rate.
The other two lines are the rates adopted in \citet{spolyar09}, for comparison.}
\label{f2}
\end{center}
\end{figure}

Suppose a ``cloud'' of DM particles contract 
from a radial position $R_{\mathrm{pre}}$ to $R_{\mathrm{new}}$, 
and the total mass distribution changes from $M_{\mathrm{pre}}(R)$ to $M_{\mathrm{new}}(R)$. 
Then the following relation holds 
during adiabatic contraction: 
\begin{equation}
	M_{\mathrm{new}}(R_{\mathrm{new}})R_{\mathrm{new}} = M_{\mathrm{pre}}(R_{\mathrm{pre}})R_{\mathrm{pre}} .
	\label{eq:Blumenthal_method}
\end{equation}
The ``new'' DM density profile can be simply calculated from the above
equation. Note that, in our case, the gas distribution also changes
because of the gas accretion onto the central primordial protostar.
We model the gas mass accretion rate in a simple
parameterized form as a function of the stellar mass. 
We use the accretion rate calculated by cosmological simulations of \citet{gao07}. 
They found a large variation of accretion rates.
In particular, the accretion rates for rotationally supported disks are 
typically smaller than for the other cases. 
As our fiducial model, we choose the accretion rate of their R5 run,
which is well described as a power-law
\begin{equation}
	\frac{{\rm d}M}{{\rm d}T}=0.18 \times M^{-0.6} \ \ \mathrm{M_{\odot} \ yrs^{-1}}.
	\label{eq:G-rate}
\end{equation} 
Hereafter we call this rate ``G-rate'' (see Figure \ref{f2}). 
The rate is for a cosmological halo forming from a very high-$\sigma$ peak,
and is slightly larger than that adopted in a previous work \citep{spolyar09}.
We also run models 
with three different accretion rates $1.0,\ 0.5,\ 0.2 \times $G-rate. 
The intermediate value, $0.5 \times $G-rate, is close to that used in \citet{spolyar09}.
We use this model to compare with their result.

The energy generation rate of DM annihilation is given by
\begin{equation}
	Q_{\mathrm{DM}} = {{\langle \sigma v \rangle {\rho}_{\chi}^2}\over{m_{\chi}}} \ \ \mathrm{GeV \ cm^{-3} \ s^{-1}}
	\label{eq:Q_DM}
\end{equation}
where $\langle \sigma v \rangle$ is the DM self-annihilation rate 
in units of ${\rm cm}^{3} \ {\rm s}^{-1}$, 
and $m_{\chi}$ is the DM particle mass in units of 
GeV = $10^9$ electron volt (e.g. \citet{bertone05}). 
Notice that effectively only the ratio, 
$\langle \sigma v \rangle/m_{\chi}$, determines
the net energy generation rate. 
We vary the DM particle mass as a model parameter as
$m_{\chi}= 1,\ 10,\ 20,\ 50,\ 100$ and $200 \ \mathrm{GeV}$,
while fixing $\langle \sigma v \rangle$ constant.
\citet{spolyar08} calculate the critical transition gas density
above which most of the DM annihilation energy is absorbed inside the core.
The critical density is smaller than the stellar density with the WIMP mass 1 GeV to 10 TeV, 
and thus we assume that the released energy from DM annihilation quickly thermalize the star.
Through the self-annihilation process, DM changes into multiple species including neutrinos, 
which escape away from the star and the cloud.
We assume that about one third of DM annihilation energy is carried away from the star, 
and thus the effective energy generation rate is $(2/3) Q_{\mathrm{DM}}$ (see, e.g. \citet{spolyar08}).

All our calculations start 
from a protostar with $M\sim10 \ \mathrm{M_{\odot}}$.
As the initial condition, we calculate an equilibrium stellar structure for this mass, 
including DM annihilation energy.
In order to apply the Blumenthal's method for the adiabatic contraction calculation,
we need the gas density profile outside the star. 
We use the result in the cosmological simulation of
\citet{gao07} which shows the gas density $\rho$ scales with distance $r$ 
as $\rho_{\rm gas} \propto r^{-2.3}$. 
Within an integration time step, the accreted gas is added  
to the stellar mass and DM annihilation energy is re-calculated. 
Because the DM density increases extremely within the central star
due to adiabatic contraction, the DM annihilation rate 
at the outside of the star is relatively small than inside.
Thus we consider the effect of DM annihilation inside the star only.
We stop the calculation when the star contracts to be sufficiently
compact and the stellar surface temperature 
reaches $10^5 \ \mathrm{K}$. The dark star 
phase ends well before entering this regime.
Afterwards the star is expected to settle quickly on the main-sequence phase, 
mainly powered by nuclear reactions. 
In this study, we do not include nuclear reactions 
for our dark star models, because the nuclear reactions are ineffective 
owing to the low temperature inside the star during the dark star
phase.

In the next section, we first show the results from
the run with a fiducial set of parameter values. 
We call this model ``base model''.
The parameters of our base model are
a power-law gas mass accretion rate $1.0 \times $G-rate, 
$m_{\chi}=100 \ \mathrm{GeV}$
and $\langle \sigma v \rangle=3 \times 10^{-26} \ \mathrm{cm^3 \ s^{-1}}$.
We will also show the results of models which adopt different values for 
these two parameters.

\section{Results}

\subsection{Base model}

\begin{figure}
\begin{center}
\includegraphics{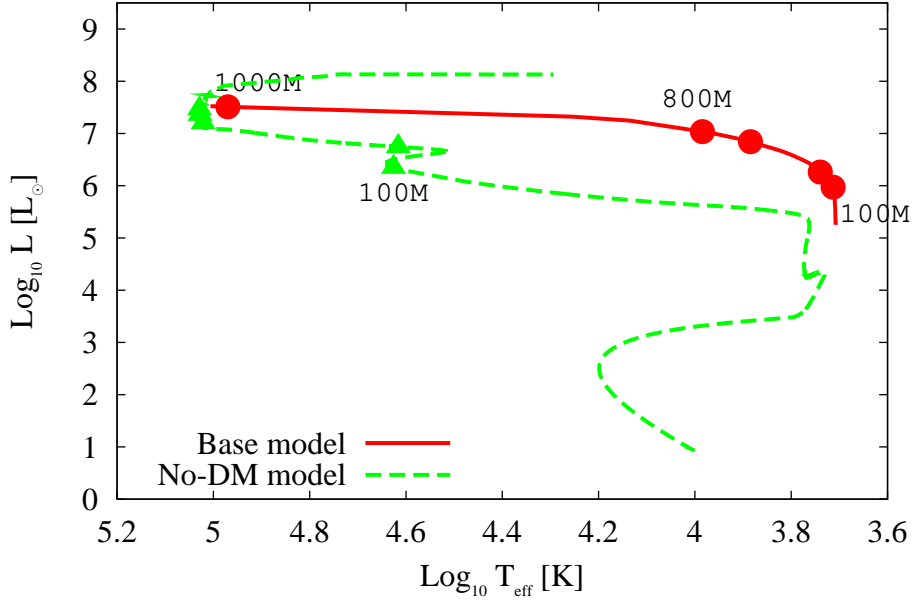}
\caption{
Hertzsprung-Russell (HR) diagram for our base model.
Symbols indicate the stellar mass of 
$M=100,\ 200,\ 600,\ 800$ and $1000 \ \mathrm{M_{\odot}}$. 
The solid line is for the dark star model (``base model''), 
whereas the dashed line is for the standard Pop III model 
(no-DM model of \citet{umeda09}).}
\label{f3}
\end{center}
\end{figure}

Figure \ref{f3} shows the evolution of the dark star
in the Hertzsprung-Russell (HR) diagram. 
The accreting star grows constantly in mass. In the HR diagram,
the star moves upward and then turns to the left 
(higher temperature). 
The solid line is for our dark star model, whereas 
the dashed line is for a standard Population III model 
(without DM annihilation energy; no-DM model, \citet{umeda09}). 
Symbols indicate the times when the stellar mass is 
$M = 100,\ 200,\ 600,\ 800$ and $1000 \ \mathrm{M_{\odot}}$. 
Clearly the dark star model moves on a very 
different path from that of the no-DM model,
showing a significant effect of the DM annihilation energy.
In the no-DM model, the star contracts gravitationally 
and becomes hot quickly until the first symbol ($100 \ \mathrm{M_{\odot}}$). 
The dark star, on the other hand, has essentially the same effective temperature 
because DM annihilation energy prevents the star from gravitational contraction.
When the no-DM star reaches  $\sim 100 \ \mathrm{M_{\odot}}$, 
it expands, and turns to the lower temperature side in the HR diagram.
There, the hydrogen burning becomes effective to supply energy 
to stop the contraction and the star expands slightly.
Subsequently the star lands on the main-sequence phase.

Contrastingly, the dark star remains ``cool'' while it 
is powered by DM annihilation. 
This is because the DM annihilation energy (Eq. \ref{eq:Q_DM}) 
is independent of the stellar temperature.
Even if the stellar interior remains at low temperatures, 
DM annihilation can produce enough energy 
to sustain the stellar structure.
As long as this condition is met, the star cannot contract
while still grows in mass by accretion.
Interestingly, by the time when the dark star reaches main-sequence, 
the stellar mass is already $\sim 1000 \ \mathrm{M_{\odot}}$, 
which is much larger than the no-DM case ($100 - 200 \ \mathrm{M_{\odot}}$).

\begin{figure}
\begin{center}
\begin{tabular}{cc}
\includegraphics[scale=.6]{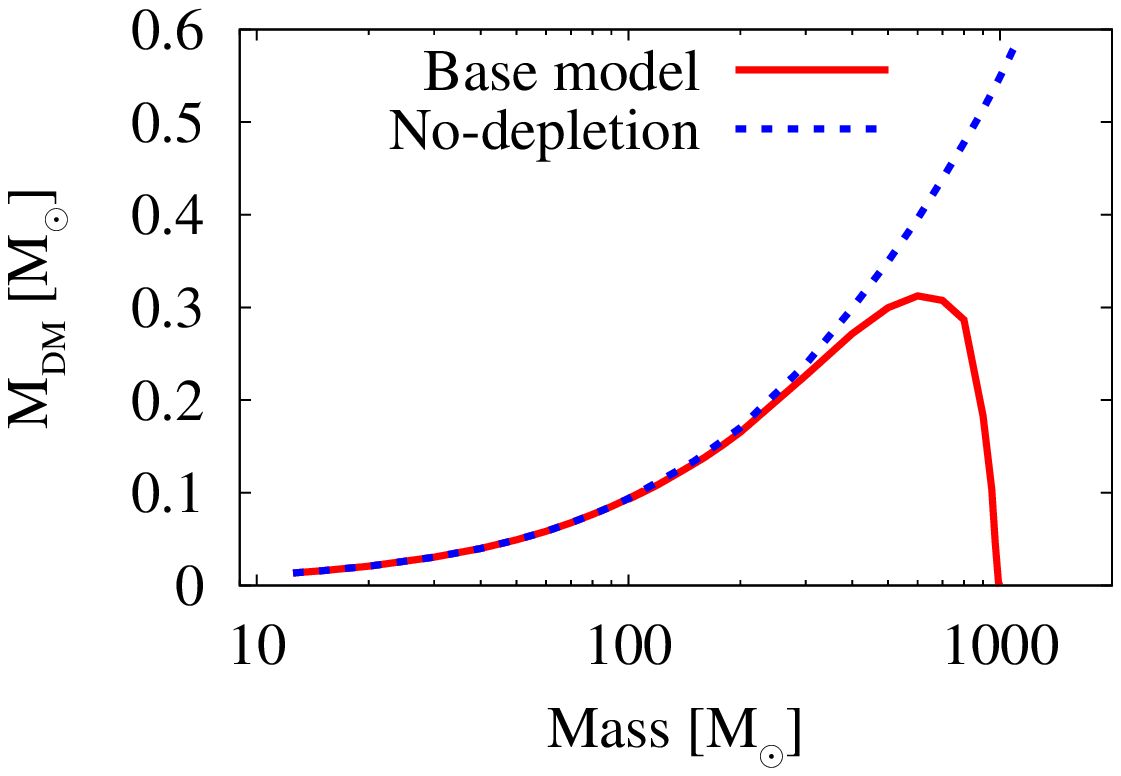} & \includegraphics[scale=.6]{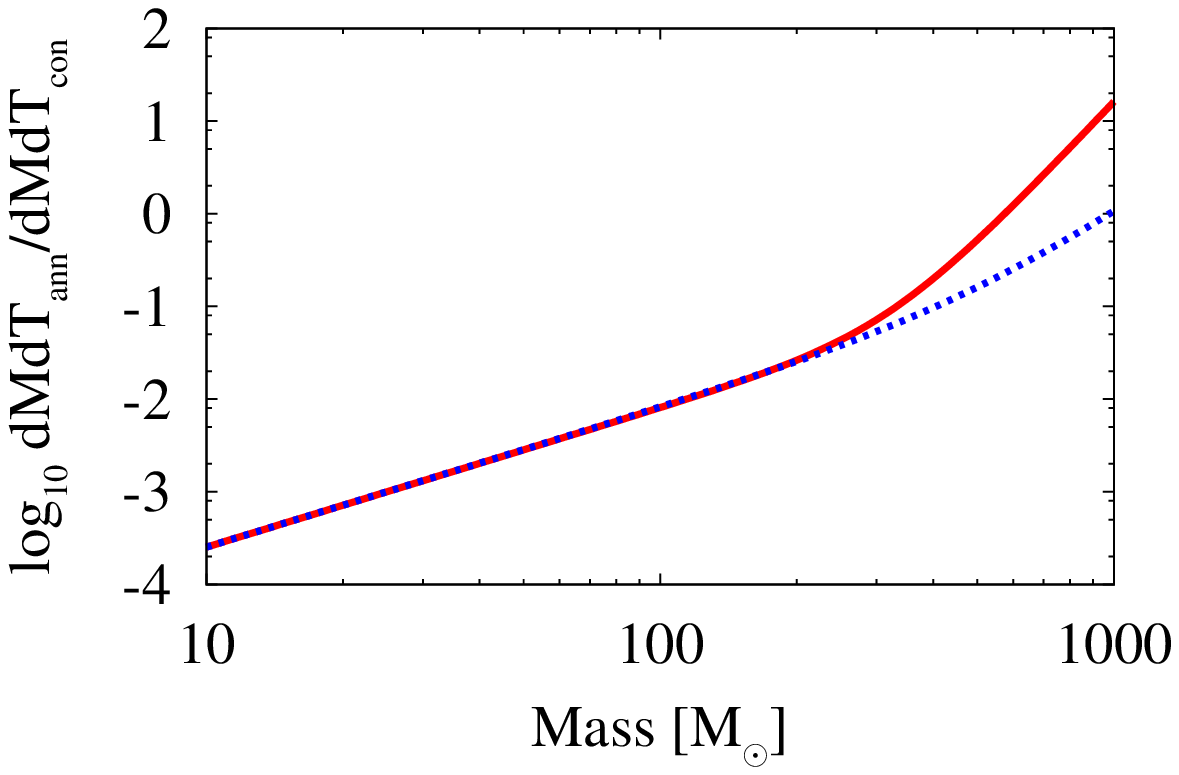}
\end{tabular}
\caption{
DM mass inside the star (left panel) and the DM consumption rate 
by the annihilation (right panel) as a function of stellar mass. 
We normalize the DM consumption rate ${{dM_{DM}}/{dt}}|_{\mathrm{annihilation}}$
by the DM mass supply by adiabatic contraction 
${{dM_{DM}}/{dt}}|_{\mathrm{contraction}}$. 
The solid line is for the dark star model (``base model''), 
whereas the dotted line is for no-depletion model.}
\label{f4}
\end{center}
\end{figure}

Figure \ref{f4} shows the evolution of 
the DM mass inside the star and its time derivative.
The left panel shows the total DM mass inside 
the star, whereas the right panel shows the ratio of 
the rate of DM annihilation in mass
to the rate of DM mass increase (within the star) owing to adiabatic contraction:
\begin{equation}
\left. \left. \frac{{\rm d}M_{\rm DM}}{dt}\right|_{\mathrm{annihilation}}
\right/
\left. \frac{{\rm d}M_{\rm DM}}{dt}\right|_{\mathrm{contraction}}.
\end{equation} 
In Figure \ref{f4}, 
we show the results for an additional model
in which the depletion of DM 
in the star is {\it not} taken into account.
Although this case might appear unrealistic, it is useful to 
see when DM depletion becomes substantial.
The solid line represents our ``base model'' 
while the dashed line shows the result of the no-depletion case. 
Up to $M \sim 200 - 300 \ \mathrm{M_{\odot}}$, 
there is almost no difference between the two cases. 
At $M > 300 M_{\odot}$, however, DM depletion becomes significant 
and then the total DM mass actually starts decreasing.
The evolution thereafter is interesting.
The DM ``fuel'' inside the star runs short to sustain the star. 
The star stops expanding, the gas density increases more efficiently,
and then DM density also increases again. 
Then DM annihilation consumes rapidly the DM fuel inside the star.
Finally, the DM annihilation energy cannot sustain the star, 
marking the end of the dark star phase. The star collapses and 
will eventually reach the main-sequence phase.

\begin{figure}
\begin{center}
\begin{tabular}{cc}
\includegraphics[scale=.6]{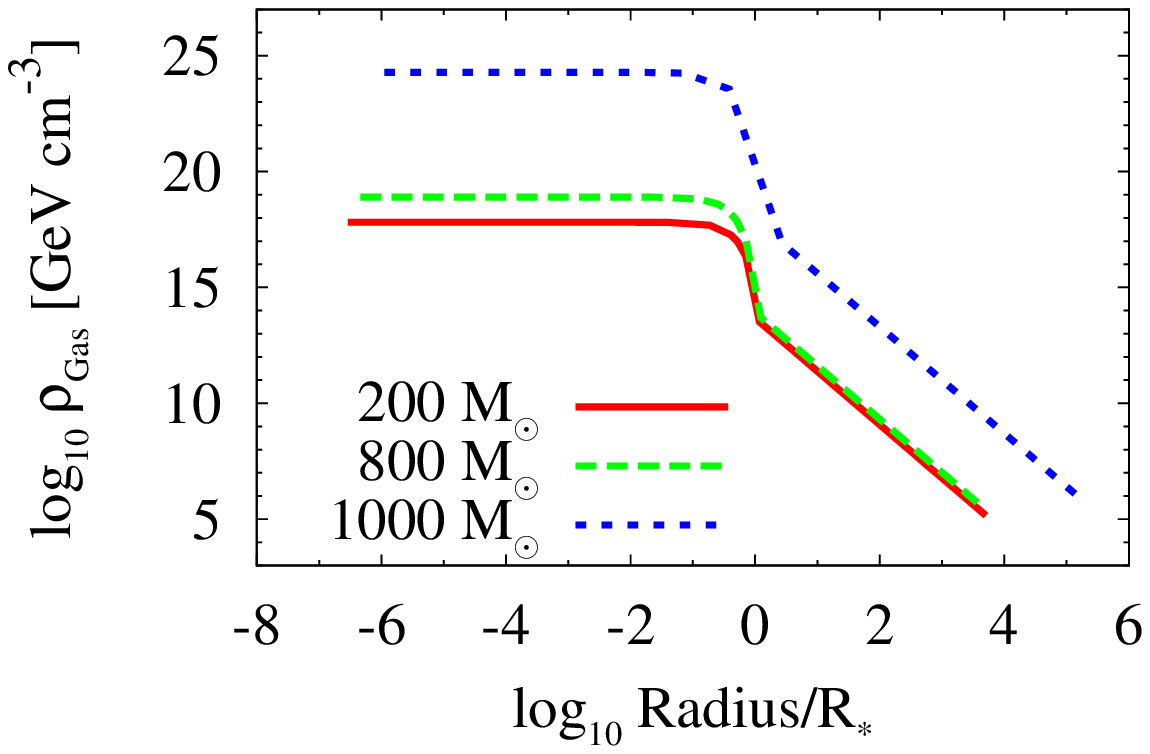} & \includegraphics[scale=.6]{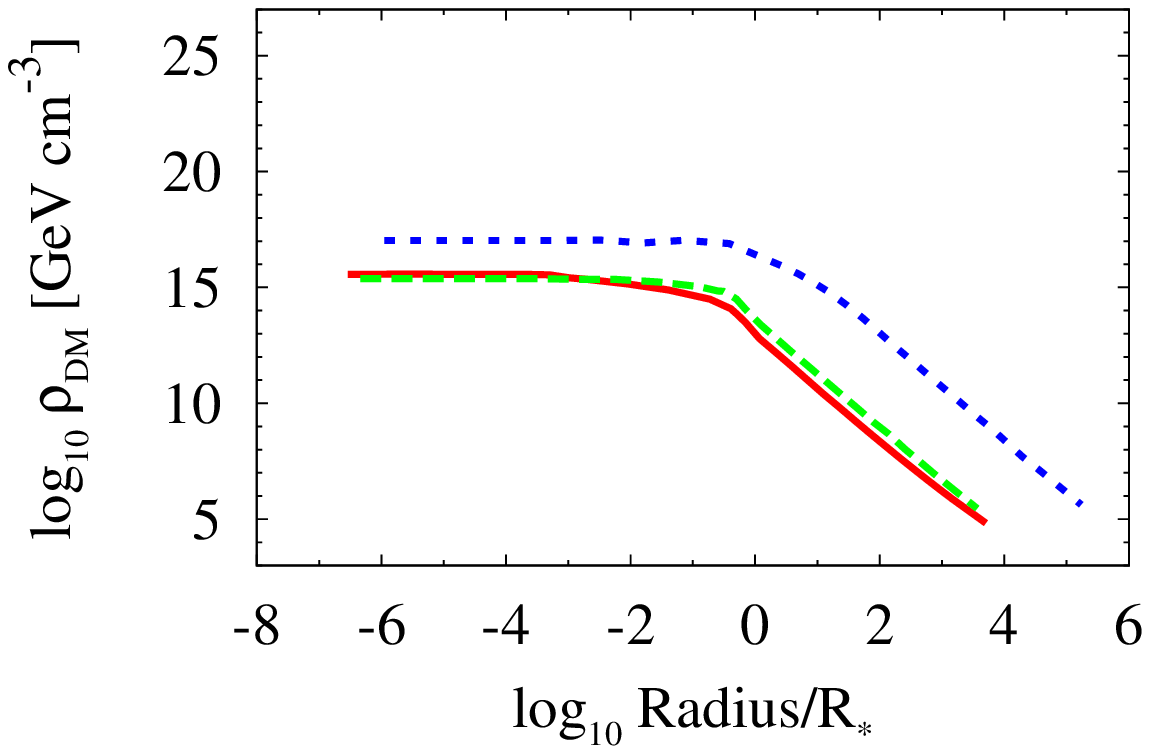} \\
\includegraphics[scale=.6]{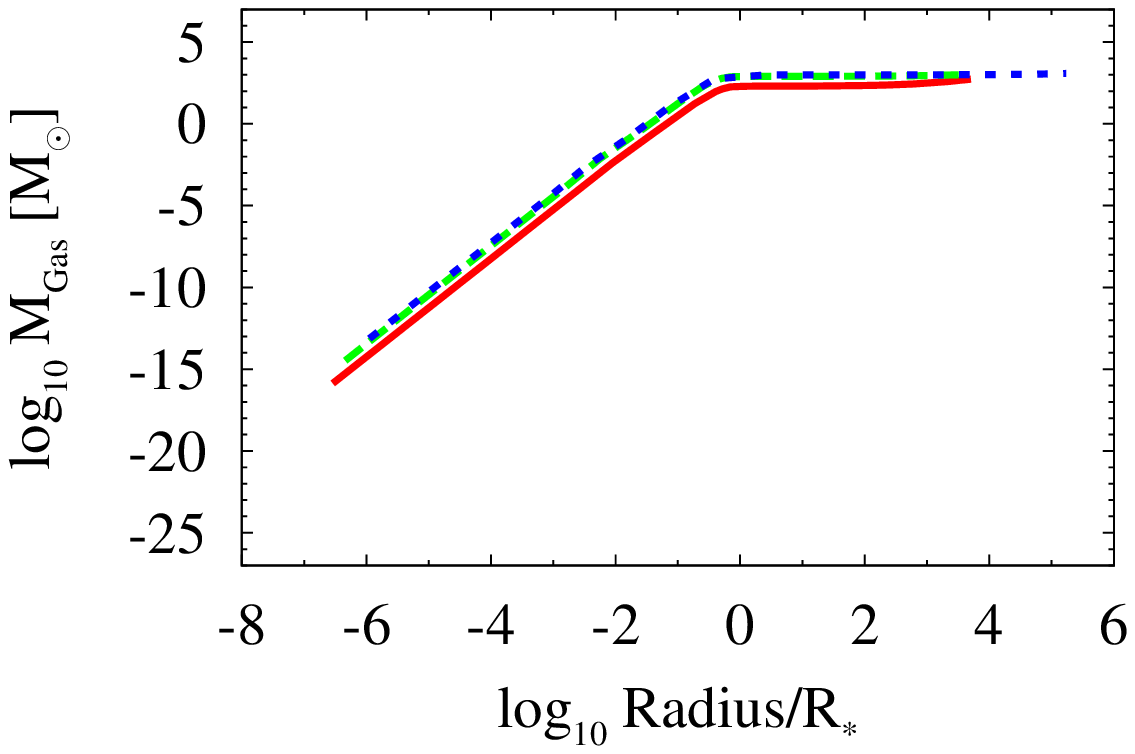} & \includegraphics[scale=.6]{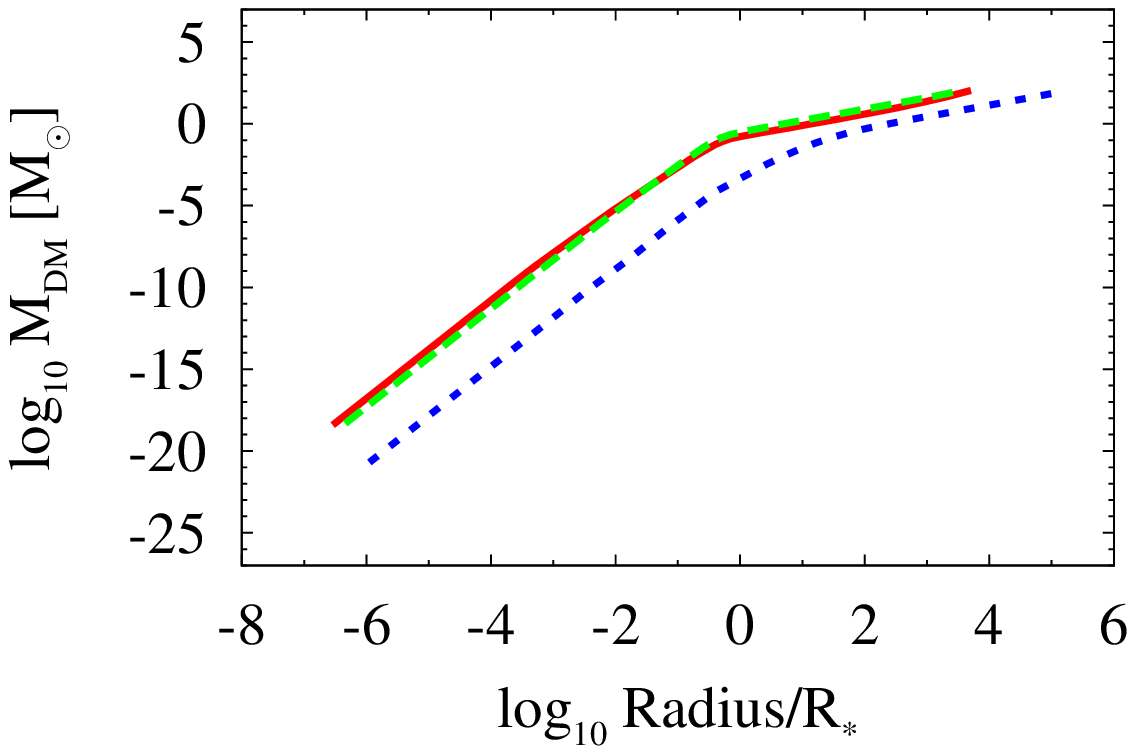} \\
\includegraphics[scale=.6]{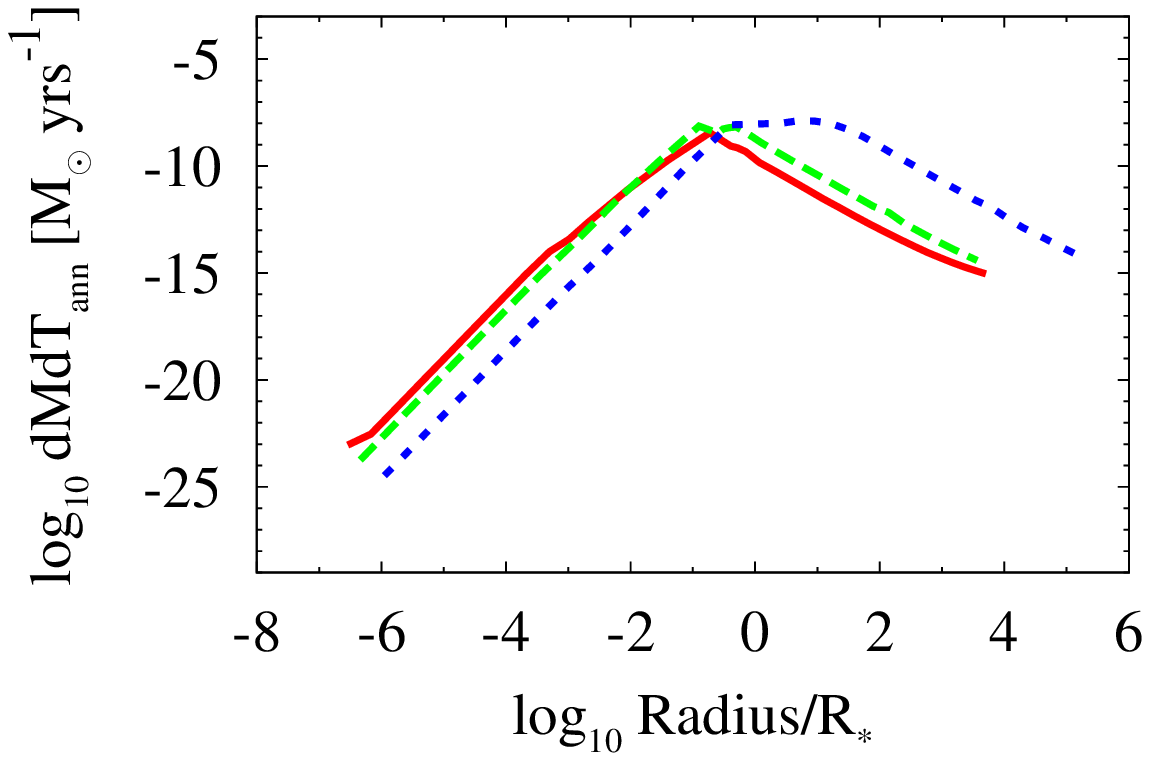} & \includegraphics[scale=.6]{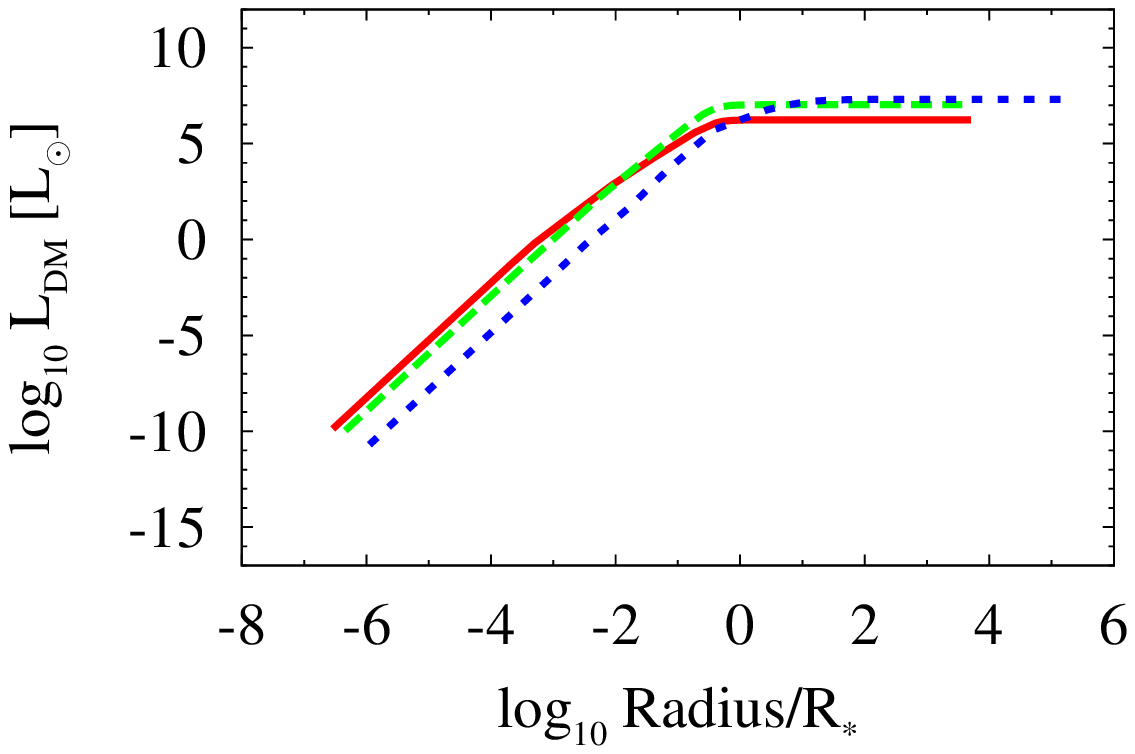}	
\end{tabular}
\caption{
Radial profiles of various quantities for our base model 
when the stellar masses are 
$M=200,\ 800$ and $1000 \ \mathrm{M_{\odot}}$. 
In all the panels, the horizontal axis shows the ratio of radial distance
normalized to the stellar radius (see Table \ref{table1}). 
Top panels show the gas and DM density profiles. The middle panels show 
the gas and DM enclosed mass. 
The bottom panels show DM annihilation rate and DM luminosity from annihilation.}
\label{f5}
\end{center}
\end{figure}

Figure \ref{f5} shows the radial profiles for
various quantities of gas ({\it left}) and DM ({\it right}) 
when the stellar mass is $M = 200$ (solid lines),
800 (dashed lines) and $1000 \ \mathrm{M_{\odot}}$ (dotted lines)
respectively. 
The horizontal axis is the radial distance from the stellar center 
divided by the stellar radius. 

The top panels show the evolution of gas density and DM density.
Between $M=200$ to $800 \ \mathrm{M_{\odot}}$, 
the star has an extended structure and the central density does
not increase much.
The DM density increases by adiabatic contraction 
but actually decreases slightly at the inner most part owing to annihilation.
At the final contraction phase ($M=800$ to $1000 \ \mathrm{M_{\odot}}$), 
both the gas density and the DM density increase substantially.

The middle panels show the evolution of enclosed mass of gas and DM. 
Although the gas mass profile stays roughly unchanged,
the DM mass decreases dramatically during the final phase
from $M=800$ to $1000 \ \mathrm{M_{\odot}}$.
Note that the horizontal axis in the plots shows a normalized radius
$R/R_{*}$. The stellar radius $R_{*}$ itself changes significantly
over the plotted range of evolutionary stages.
Because the dark star phase ends with the runaway burning of DM inside 
of the star, the total amount of DM when $M=1000 \ \mathrm{M_{\odot}}$
is already very small.

The bottom-left panel shows the DM annihilation rate and 
the bottom-right panel shows the total luminosity generated 
by the DM annihilation within the radius.
These panels show the energy generating efficiency from the DM annihilation.
Again, we see that the annihilation rate is low inside the star at
the final stage $M=1000 \ \mathrm{M_{\odot}}$ and
the enclosed DM luminosity at the stellar surface 
(at $\mathrm{log_{10} Radius/R_* = 0}$) is smaller than in the earlier phases.
The stellar mass increases by gas accretion 
but the DM energy supply decreases;
this causes the star to contract.

\begin{figure}
\begin{center}
\includegraphics[scale=1.4]{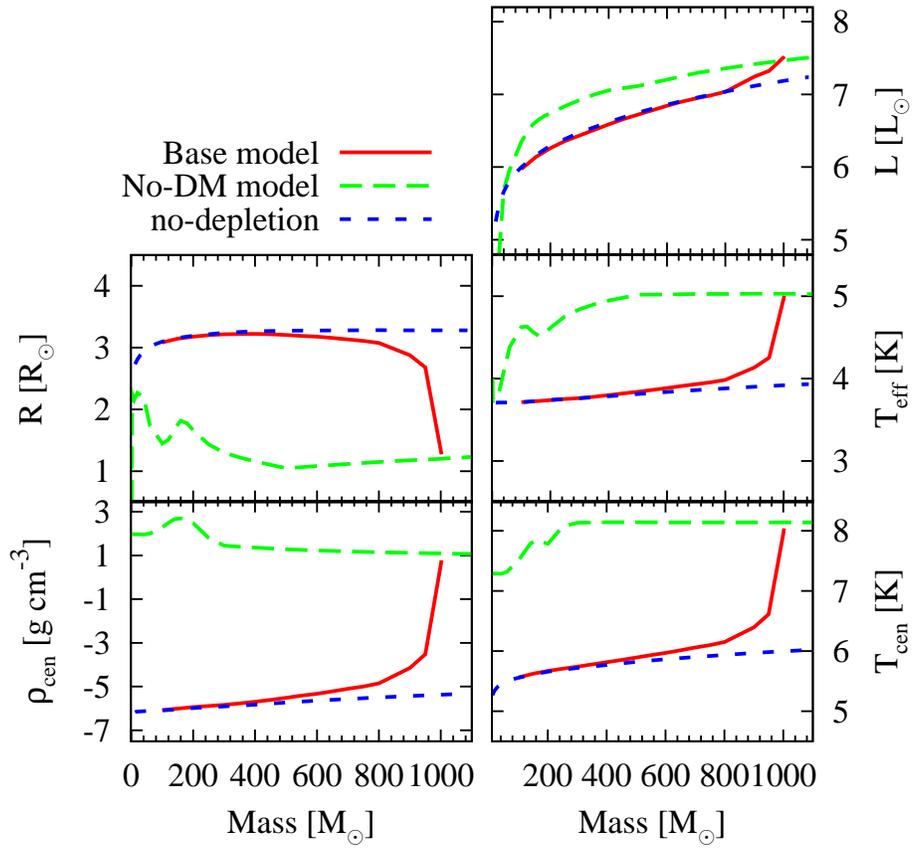}
\caption{
Evolution of basic stellar quantities.
The horizontal axis indicates the stellar mass.
The vertical axis shows, respectively, 
the stellar radius, the central gas density, the luminosity
, the surface temperature and the central temperature 
in the logarithm scales. 		
The solid line is for the base model, the dashed line is for no-DM model 
and the dotted line is for no-depletion model.}
\label{f6}
\end{center}
\end{figure}

Figure \ref{f6} shows the evolution of some 
basic stellar quantities which characterize the dark star.
We compare the results of our base model (solid lines)
with those of no-DM model (dashed lines) and of no-depletion model
(dotted lines).
Until the star grows to $M \sim 200 - 300 \ \mathrm{M_{\odot}}$, 
the base model and the no-depletion model appear very similar,
showing again that DM depletion is negligible in early phases.
After the star grows to $\sim 300 \mathrm{M_{\odot}}$, 
we see small but appreciable differences
between the base model and the no-depletion model.
As has been shown in Figure \ref{f4},
DM is consumed by annihilation inside the star whereas
the supply by adiabatic contraction is slow
owing to the small gas mass accretion rate.
As the star becomes more massive, it needs more DM 
to produce necessary energy to sustain gravitational equilibrium. 
When the DM supply becomes insufficient, this quasi-stable dark star phase 
cannot be sustained.
It occurs at $M \sim 600 \ \mathrm{M_{\odot}}$ for the base model.
The stellar properties, however, do not change immediately 
until up to $M \sim 900 \ \mathrm{M_{\odot}}$. 
At around $M \ge 900 \ \mathrm{M_{\odot}}$, the DM inside the star 
burns out rapidly, the star begins to collapse, and
the central temperature increases rapidly.
Finally the central temperature reaches $10^7 - 10^8 \ \mathrm{K}$,
the nuclear burning will soon start, and the star will eventually 
land on the main-sequence. 

Table \ref{table1} summarizes the basic stellar properties 
at several characteristic phases. 
Note that the elapsed time difference (the right column) is large
between $800$ and $1000 \ \mathrm{M_{\odot}}$. 
The dark star phase continues for about $\sim0.2 \ \mathrm{Myr}$.
Throughout the run, the DM mass is very small compared with 
the stellar gas mass. 
Dark stars can be sustained by the DM of only $0.1 \%$ of their total mass 
$M_{*}$. 

\subsection{The effect of gas mass accretion rate}

We now explore a few parameter spaces.
Our dark star model has essentially two parameters, 
the gas mass accretion rate and the DM particle mass. 
First, we examine the effect of the gas mass accretion rate. 
We adopt three accretion rates, 
$1.0,\ 0.5,$ and $0.2 \times $G-rate (Figure \ref{f2}).
``G-rate'' is the gas accretion rate in our base model (Eq. \ref{eq:G-rate}). 
In general, less DM is attracted toward the star
for smaller gas accretion rates.
We consider only smaller accretion rates than G-rate
because it seems reasonable to assume that various proto-stellar
feedback effects, such as ionizing radiation from the star,
can only reduce the gas mass accretion rate.

\begin{figure}
\begin{center}
\includegraphics{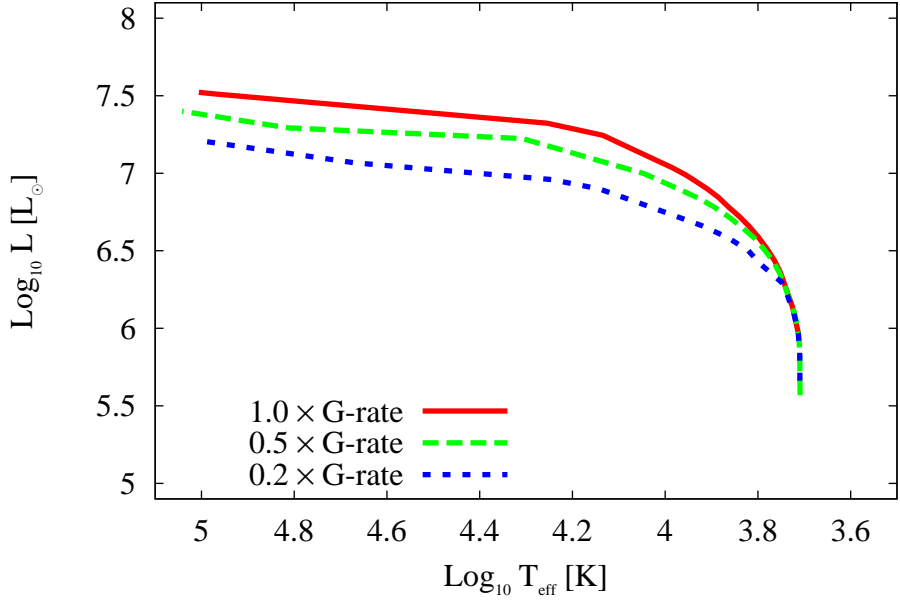}
\caption{
HR diagram for models with three accretion rates,
$1.0,\ 0.5$ and $0.2 \times $ G-rate, 
shown in Figure \ref{f2}.}
\label{f7}
\end{center}
\end{figure}

Figure \ref{f7} shows the stellar evolution for the three models 
with different accretion rates in the HR diagram. 
Initially, the stars are puffy, cool and evolve along virtually the same path. 
Tables \ref{table2} and \ref{table3} show the basic 
stellar properties
when the models with the reduced accretion rates reach the end of the stable phase 
where $T_{\mathrm{eff}} = 10^4 \ \mathrm{K}$, and where
$T_{\mathrm{eff}} = 10^5 \ \mathrm{K}$, respectively. 
The latter phase is the end of our calculations.
As is expected naively, 
the final stellar mass is smaller for lower gas accretion rates,  
but the period of the dark star phase is actually longer.
In the reduced accretion models, the star needs a longer time to reach the same stellar mass.
Then the DM density inside the star is {\rm smaller} because the gas 
contraction is slow.
Consequently, the star with a smaller gas accretion rate ends the dark star phase and 
begins to contract when its mass is smaller than in the base model.

\begin{figure}
\begin{center}
\begin{tabular}{cc}
\includegraphics[scale=.6]{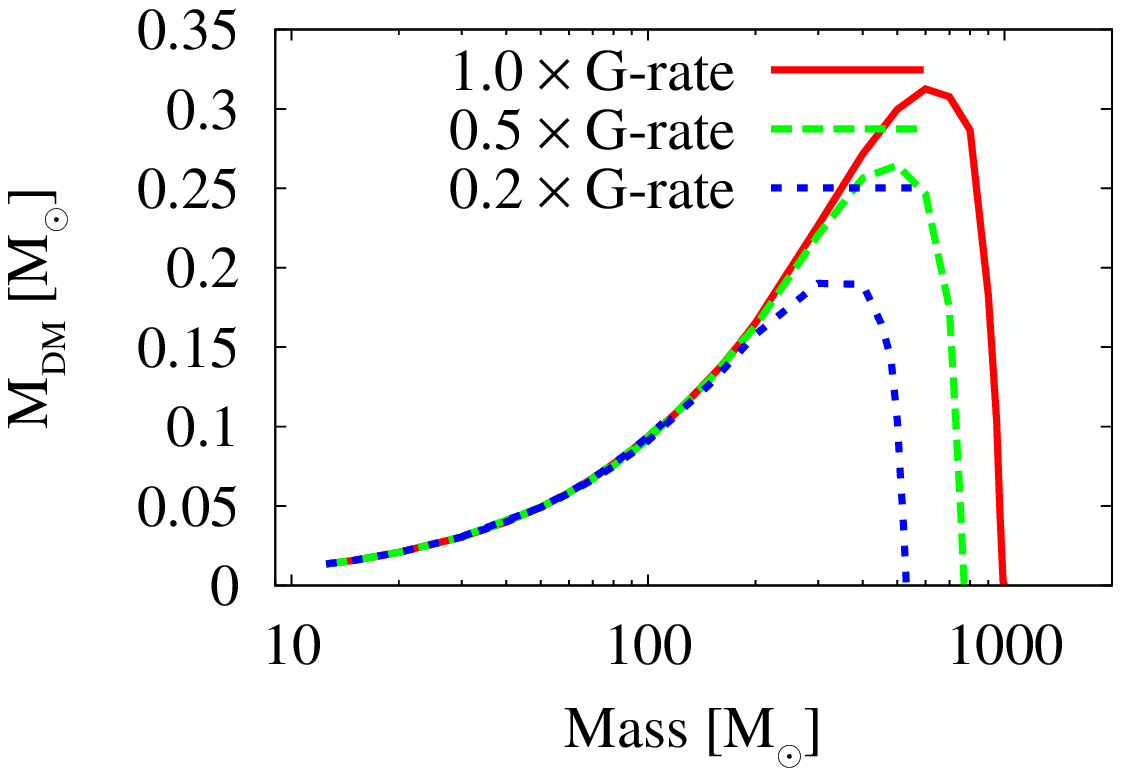} & \includegraphics[scale=.6]{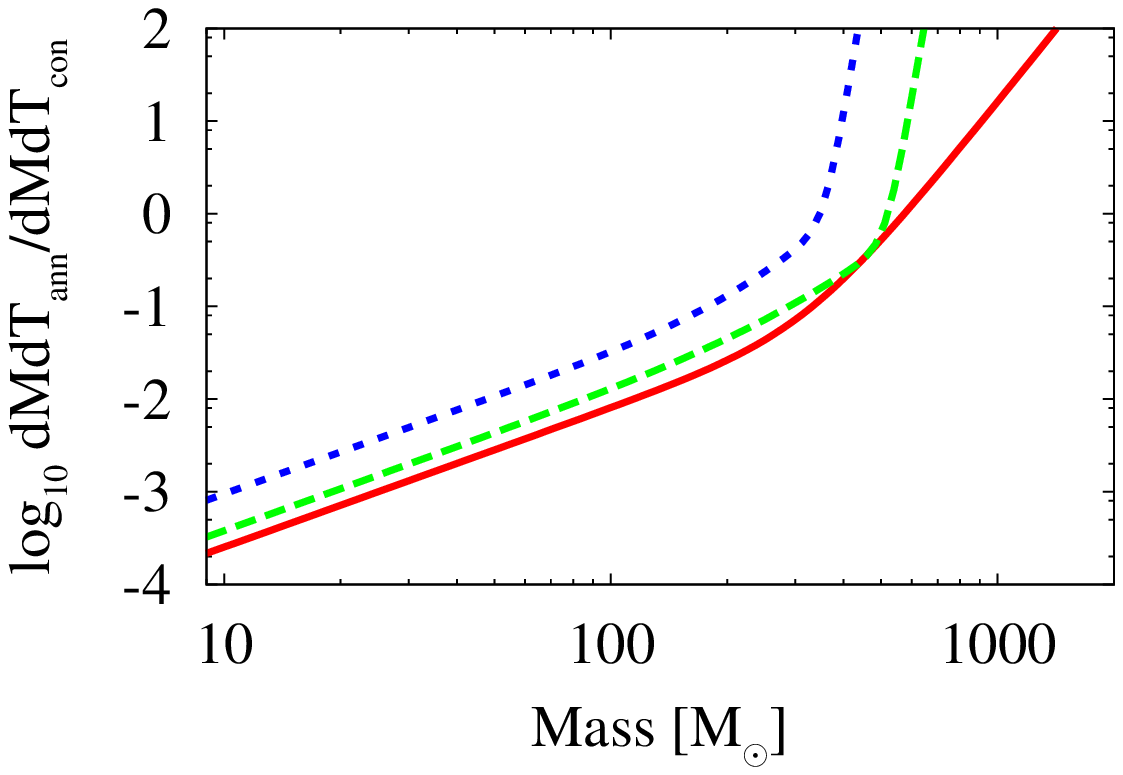}
\end{tabular}
\caption{
DM mass and the DM consumption rate by the annihilation as a function of stellar mass
for three accretion rates models, as in Figure \ref{f4}.}
\label{f8}
\end{center}
\end{figure}

\begin{figure}
\begin{center}
\includegraphics[scale=1.4]{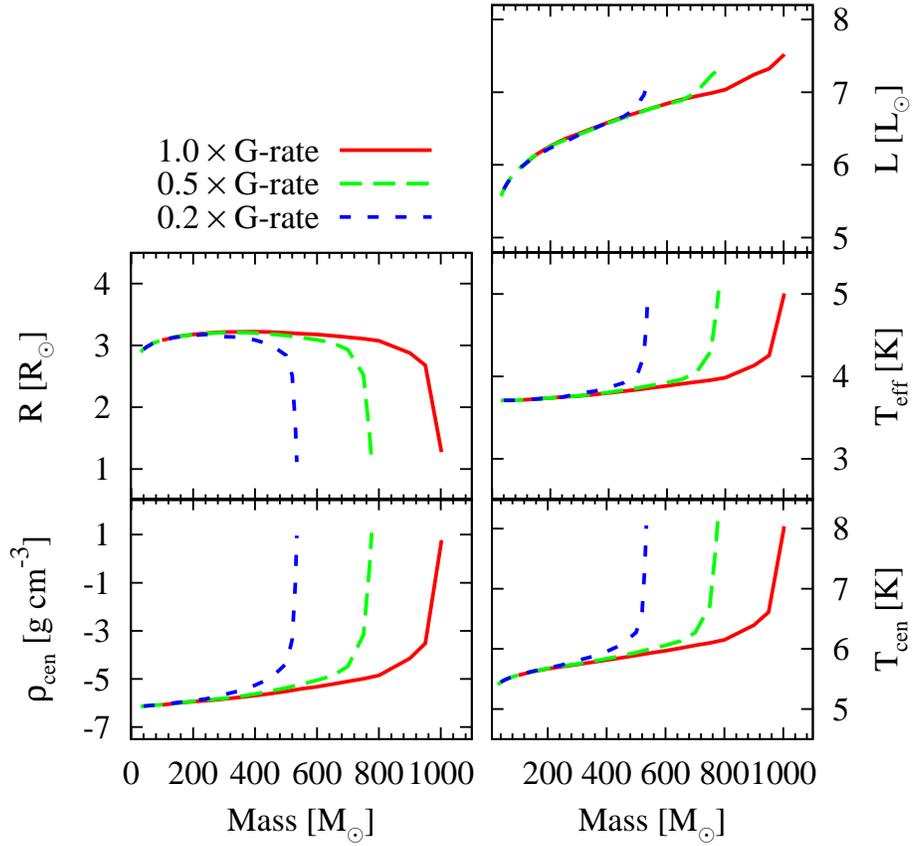}
\caption{
Evolution of basic stellar quantities for dark star models with reduced accretion rates. 
All vertical axes are plotted in logarithm scale.
The plotted quantities are the same as in Figure \ref{f6}.}
\label{f9}
\end{center}
\end{figure}

Figures \ref{f8} and \ref{f9} show clearly the effect of reducing
the accretion rate.
There are essentially no differences among the three models
when the stellar mass is small. 
The DM fuel inside the star increases very similarly (Fig. \ref{f8}a) 
because the rate of DM decrease to DM increase is very small (Fig. \ref{f8}b).
For the smallest accretion rate, the dark star ends
when the stellar mass reaches $\sim 500 \ \mathrm{M_{\odot}}$.
Overall, the lower the accretion rate is, the smaller the DM mass is inside the star. 
The dark star phase lasts longer but ends at lower masses.

\subsection{The effect of DM particle mass}

Finally, we study models with different DM particle masses. 
We adopt the base accretion rate of $1.0 \times $G-rate 
which is larger than those adopted in \citet{spolyar09} (see Figure \ref{f2}).
so our results are slightly different from theirs,
especially for a small DM particle mass case. 
We discuss this issue later in this section.

\begin{figure}
\begin{center}
\includegraphics{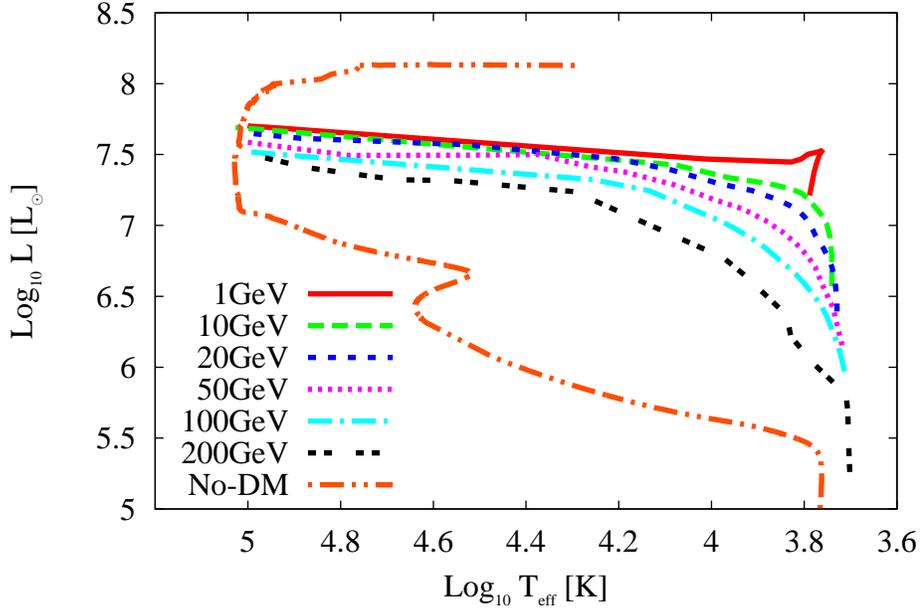}
\caption{
HR diagram for models with variation of DM particle masses.
Each lines show models with DM particle masses 
$m_{\chi}= 1,\ 10,\ 20,\ 50,\ 100, \ 200 \ \mathrm{GeV}$ and 
the no-DM model (the same as in Figure \ref{f3}), respectively.}
\label{f10}
\end{center}
\end{figure}

\begin{figure}
\begin{center}
\begin{tabular}{cc}
\includegraphics[scale=.6]{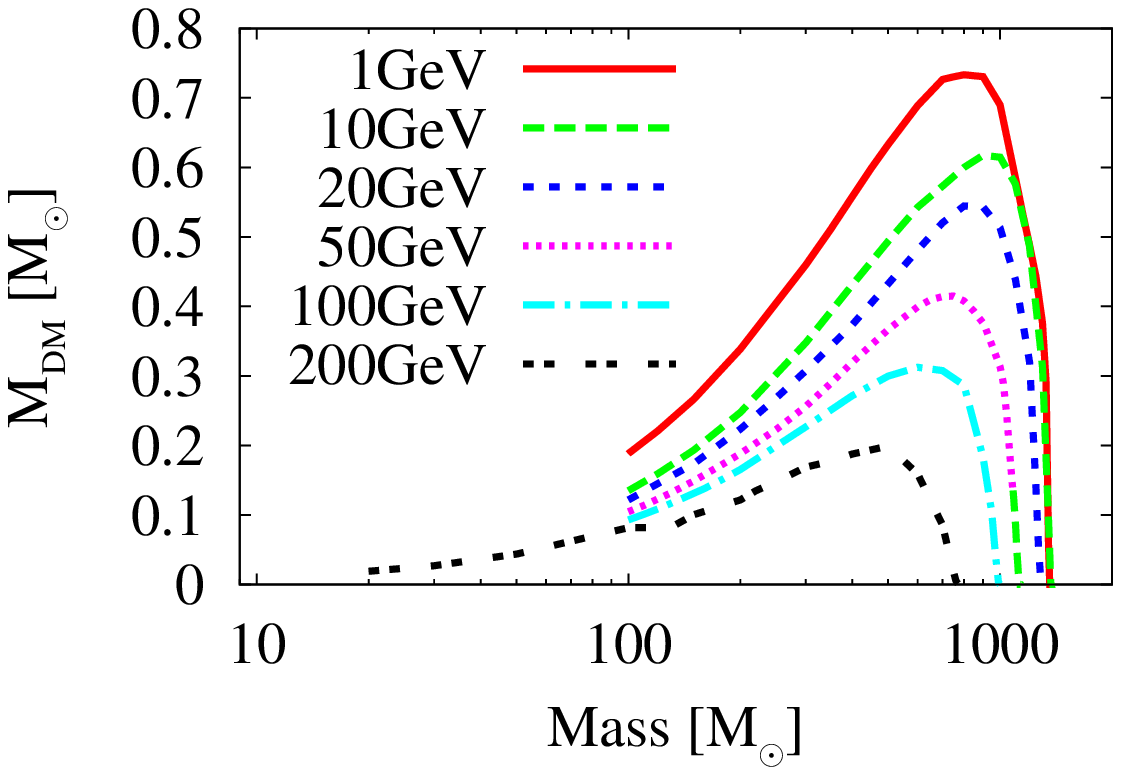} & \includegraphics[scale=.6]{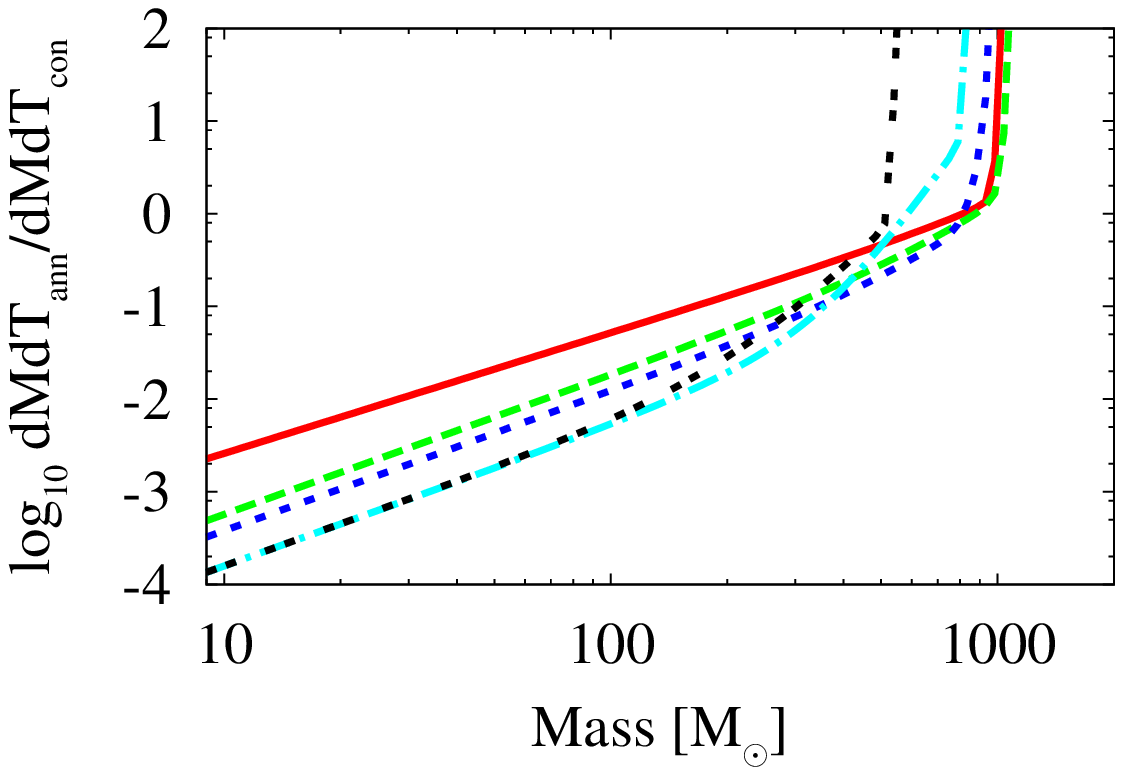}
\end{tabular}
\caption{
DM mass and the DM consumption rate by the annihilation for the variation of DM particle masses,
as the same plot of Figure \ref{f4}.}
\label{f11}
\end{center}
\end{figure}

\begin{figure}
\begin{center}
\includegraphics[scale=1.4]{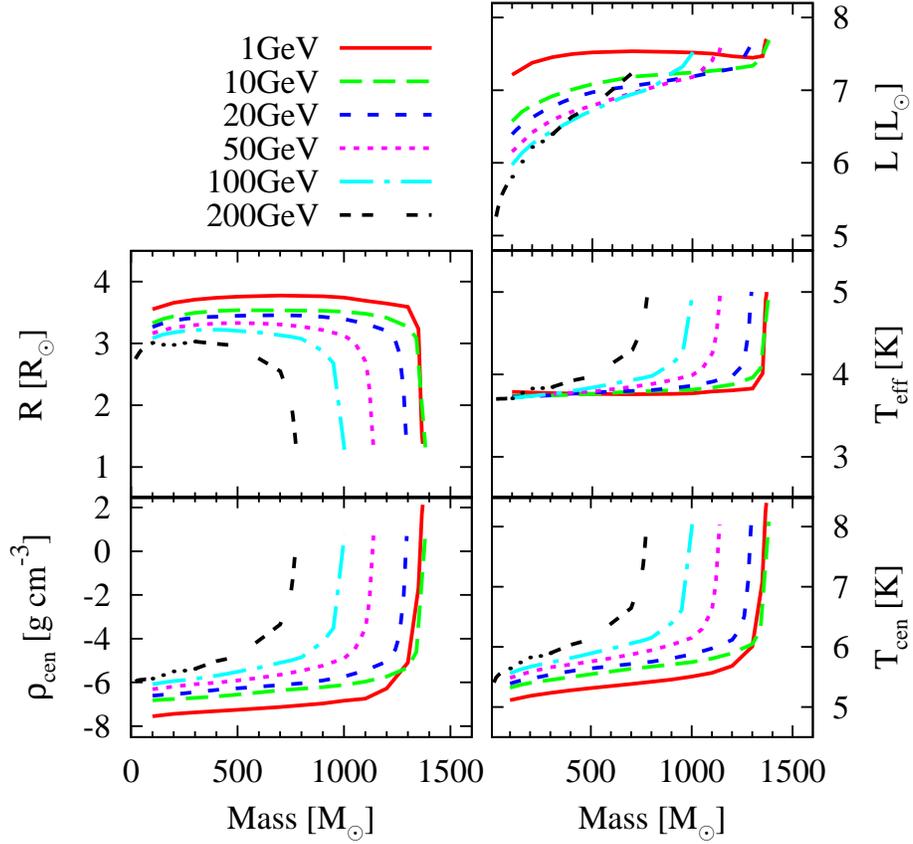}
\caption{
Evolution of stellar values for dark star models with parameters of DM particle masses.		
All vertical axes are plotted in logarithm scale.
As for Figure \ref{f6}, but for different DM particle masses.}
\label{f12}
\end{center}
\end{figure}

We run six models with
$m_{\chi}= 1,\ 10,\ 20,\ 50,\ 100$ and $200 \ \mathrm{GeV}$. 
Note that, from Eq. \ref{eq:Q_DM}, the DM annihilation energy is 
inversely proportional to $m_{\chi}$. 
Thus the characteristic features of dark stars appear strongly 
in runs with small DM masses. 
Figures \ref{f10}, \ref{f11} and \ref{f12} 
show the results for the five models.
For smaller DM particle masses, the DM annihilation rate is large,
and thus the star expands more and the temperature and density in the 
star are lower. 
The period of the dark star phase is also prolonged for smaller $m_{\chi}$, 
and the final stellar mass gets larger than our base model.

Tables \ref{table4} and \ref{table5} list
the basic properties of the stars when the surface temperature $T_{\mathrm{eff}}$ reaches 
$10^4 \ \mathrm{K}$ and $10^5 \ \mathrm{K}$, respectively. 
The final stellar mass $M_{\ast}$ differs by about a factor of two
among the five models; 
$M_{\ast}= 776$ to $1370 \ \mathrm{GeV}$ in Table \ref{table5}. 
The overall trend is as describe above, but
the lightest particle model is worth describing more in detail. 
When the star enters main-sequence, 
the stellar mass in $m_{\chi}=1 \ \mathrm{GeV}$ model 
is similar to that of $10 \ \mathrm{GeV}$ model, 
although it has 10 times larger DM energy generation rate. 
The reason is that 
the DM burning rate is too large in $m_{\chi}=1 \ \mathrm{GeV}$ model. 
When the DM consumption rate exceeds the DM supply by adiabatic contraction
(see discussion in Sec. 3.1), the star runs out of the DM fuel and
begins to contract. The DM consumption becomes even more efficient then. 
This final ``runaway'' phase takes place faster in $1 \ \mathrm{GeV}$ model 
than in $10 \ \mathrm{GeV}$ model. Consequently, the star lands on the main-sequence
early in $1 \ \mathrm{GeV}$ model (see the last column of Table 
\ref{table5} which gives the elapsed time).

\begin{figure}
\begin{center}
\begin{tabular}{cc}
\includegraphics[scale=.6]{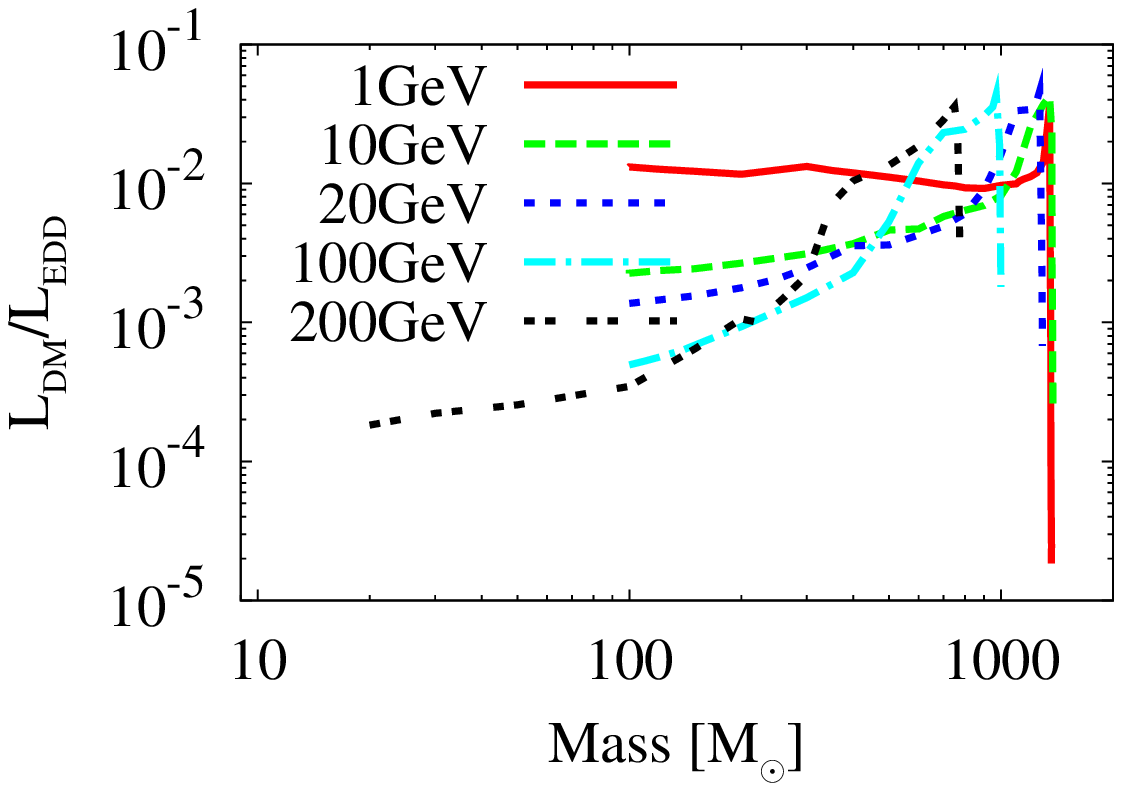} & \includegraphics[scale=.6]{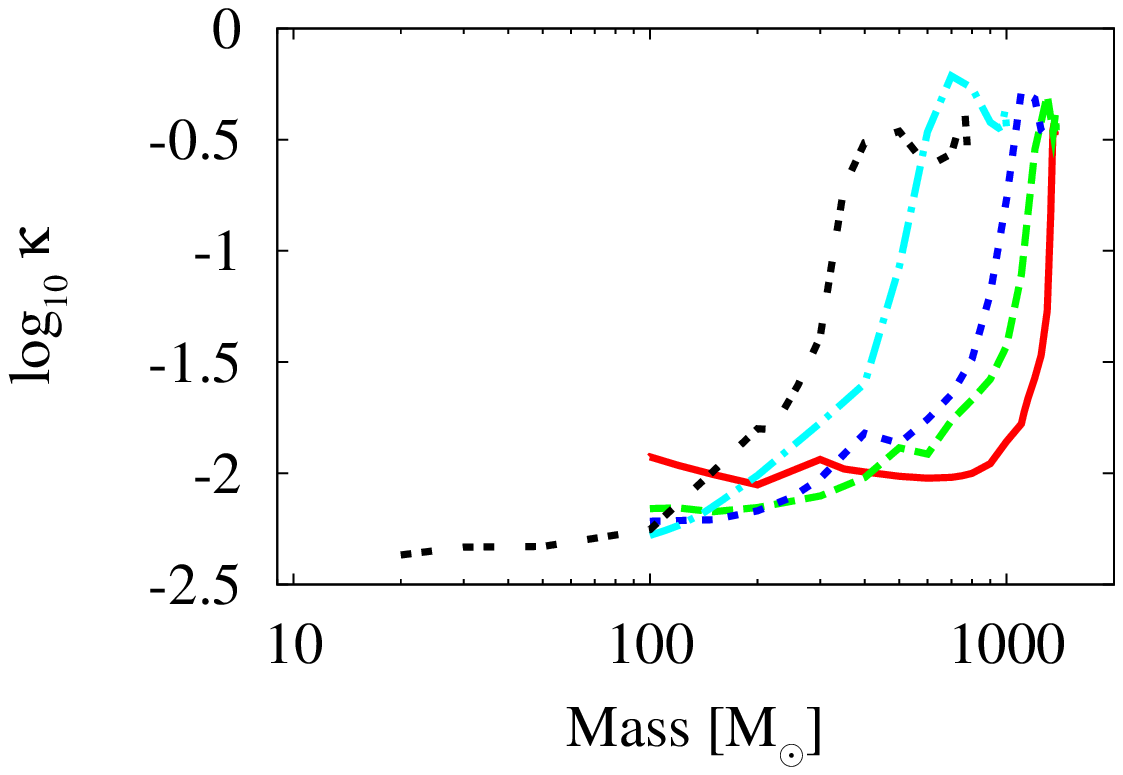}
\end{tabular}
\caption{
We plot the ratio of DM annihilation luminosity to Eddington luminosity (left) and 
the opacity at photosphere as a function of stellar mass (right)
for different DM particle masses.
In all the models, the luminosity ratio stays less than unity.
}
\label{f13}
\end{center}
\end{figure}

Interestingly, the $m_{\chi} = 1 \ \mathrm{GeV}$ model 
predicts the most luminous star, reaching $L \sim$ some $10^7 \ \mathrm{L_{\odot}}$ 
at the peak (Fig. \ref{f12}). 
The extremely high luminosity is, however, still smaller than the upper-limit 
given by the Eddington Luminosity $L_{\mathrm{EDD}}$:
\begin{equation}
	L_{\mathrm{EDD}} = {{4 \pi c G M_*}\over{\kappa}}
	\label{eq:L_EDD}
\end{equation}
where the opacity $\kappa$ determines essentially the critical luminosity.
Figure \ref{f13} shows the ratio of the energy generation rate of DM annihilation
(``luminosity'') to the Eddington luminosity for our five models with 
different DM particle masses. 
The ratio stays
always less than unity, even for $1 \ \mathrm{GeV}$ model.
During the dark star phase ($T_{\mathrm{eff}} < 10,000 \mathrm{K}$), 
the opacity $\kappa$ at lower temperatures
is mainly contributed by $H^-$ ion which has a small value. 

\section{Discussion}

\subsection{Dark Star Models}

We compare our results with those of previous works.
Unlike previous studies (e.g., \citet{spolyar09}),
we solve full radiative transfer and follow self-consistently
the stellar structure and its evolution.
A direct comparison can be made with the results of Table 2 in \citet{spolyar09} (hereafter ``SP09''),
by using the result of our $0.5\times $G-rate model 
(see Sec. 3.2; hereafter ``05G''). 
The other model parameters are set to be the same, 
$m_{\chi}=100 \ \mathrm{GeV}$
and $\langle \sigma v \rangle=3 \times 10^{-26} \ \mathrm{cm^3 \ s^{-1}}$, 
except for the gas mass accretion rate.
The gas mass accretion rates in these two models (Figure \ref{f2})
are almost the same at 
$M_* > 100 \ \mathrm{M_{\odot}}$, which is the main period of the dark star phase.
The results of 05G and SP09 are roughly similar;
in stellar radius, luminosity and other quantities, 
the differences are less than $20 \%$ 
at $M_* = 106 \ \mathrm{M_{\odot}}$ 
(the first line in SP09), 
and remain within a factor of two at later stages.
During calculation, 05G model has a slightly expanded structure 
and is somewhat cooler than SP09 model.
On the other hand, the difference of the central temperature is relatively 
large between the two models.
The value in 05G becomes lower than in SP09 especially at the final stages; 
at $M_* = 716,\ 756,\ 779\ \mathrm{M_{\odot}}$, 
$T_{\mathrm{cen}} = 2.4,\ 9.3,\ 132\ \mathrm{(10^6) K}$ in 05G 
whereas $T_{\mathrm{cen}} = 15,\ 78,\ 280\ \mathrm{(10^6) K}$ in SP09.
Because of the lower central temperature, 
nuclear burning is unimportant in our calculations.

\subsection{Possible Range of WIMP DM mass}

\begin{figure}
\begin{center}
\begin{tabular}{cc}
\includegraphics[scale=.6]{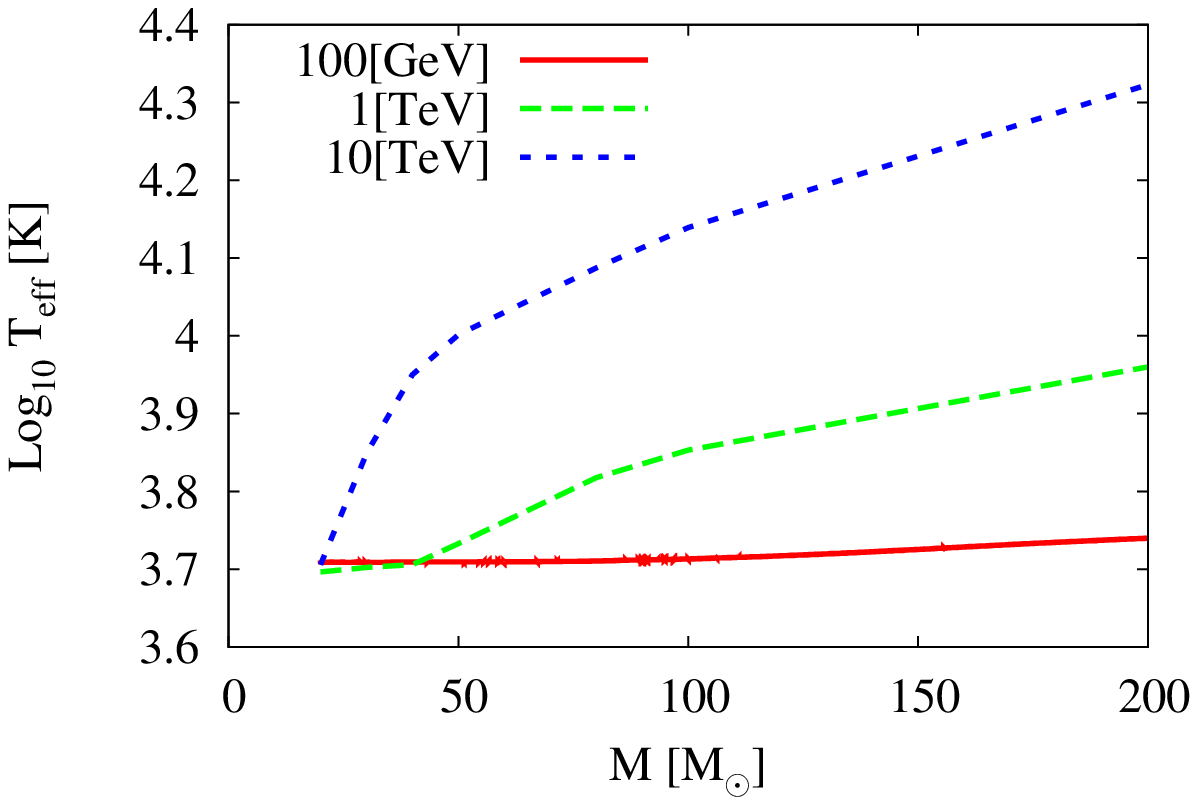} & \includegraphics[scale=.6]{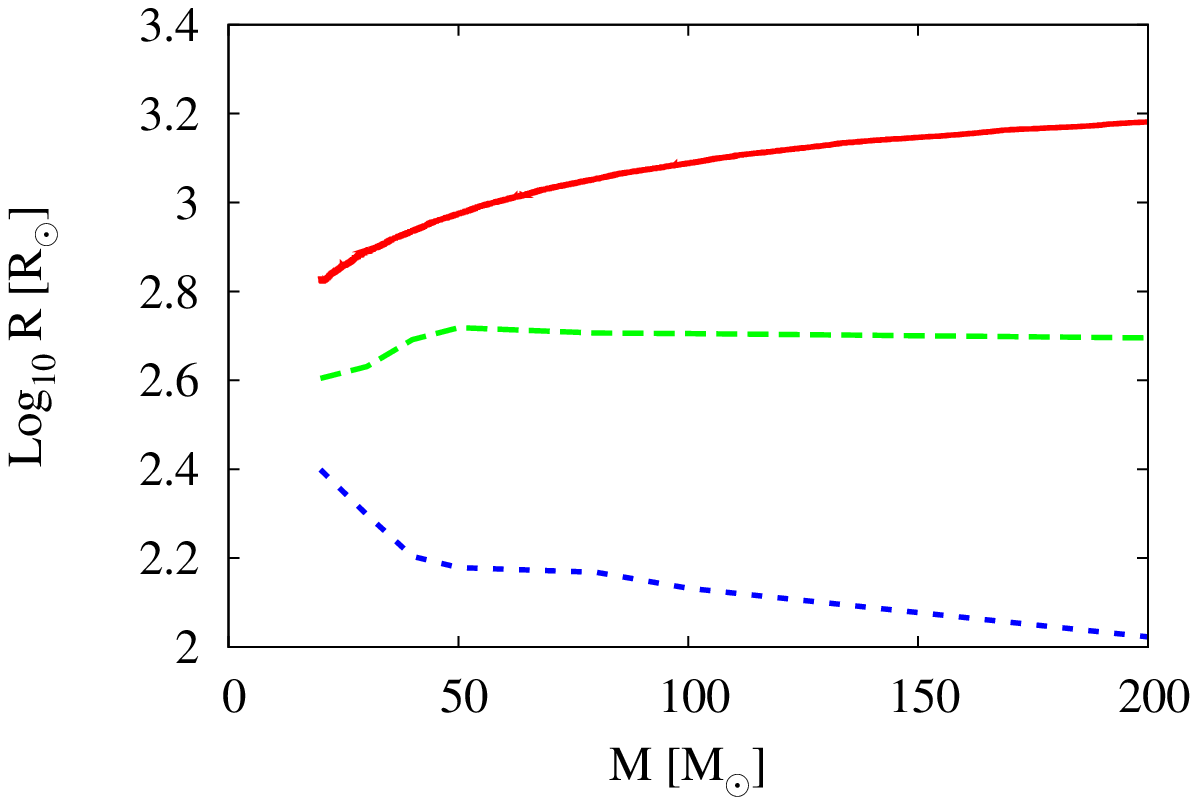}
\end{tabular}
\caption{
Stellar values of no-depletion models with large WIMP mass 
$m_{\chi}= 100 \ \mathrm{GeV}, \ 1 $ and $ 10 \ \mathrm{TeV}$.
Left panel shows the evolutions of the stellar surface temperature and right panel shows ones of the stellar radius.
All models are on the dark star phases at first.
}
\label{f14}
\end{center}
\end{figure}

The WIMP mass significantly affect the duration of the dark star phase because the
DM annihilation energy is inversely proportional to the WIMP mass $m_{\chi}$, as can be seen in Eq. \ref{eq:Q_DM}.
We have already shown the effect in Section 3.3 for a typical DM mass range, from 1 to 200 GeV. 
However, in supersymmetric particle physics models, the WIMP DM mass may be 
10 TeV or even greater.
To estimate the evolutionary path of dark star models with heavier WIMPs,
we have run our no-depletion model for two larger WIMP masses, $m_{\chi}= 1 $ and $ 10 \ \mathrm{TeV}$.
Figure \ref{f14}a and \ref{f14}b show the result.
At the beginning of the calculation with $M \simeq 15 M_{\odot}$, all the models are in the dark star phase
and on the Hayashi-line in the H-R diagram. 
The left panel of Fig. \ref{f14} shows that, at this phase, 
all models have $T_{\mathrm{eff}} \simeq 10^{3.7} \ \mathrm{K}$.
For $m_{\chi}$ = 10 TeV, the Pop.III star leaves the dark star phase early. 
Because annihilation rate is small for a large WIMP mass, the stellar mass at the end of the dark star phase remains small.

All the dark star models are on $T_{\mathrm{eff}} \simeq 10^{3.7} \ \mathrm{K}$ line at first (see also Fig. \ref{f10}) 
and leave the line gradually as mass-accretion continues. 
To quantify the evolution, we define two characteristic stellar masses, 
$M_{3.8}$ and $M_{4.0}$, the stellar masses at which the surface temperature reaches 
$T_{\mathrm{eff}} \simeq 10^{3.8}$ and $10^{4.0} \ \mathrm{K}$. 
The former indicate the phase where the star deviates from the Hayashi-line. 
The later indicates when a star ends the stable dark star phase. The result is, 
$M_{3.8} = (1145, \ 821, \ 402, \ 74, \ 26) \ \mathrm{M_{\odot}}$ and 
$M_{4.0} = (1349, \ 1325, \ 821, \ 270, \ 50) \ \mathrm{M_{\odot}}$ for 
$m_{\chi}= (1, \ 10, \ 100 \ \mathrm{GeV}, \ 1, \ 10 \ \mathrm{TeV})$, respectively.
We conclude that Pop.III stars with a large WIMP mass up to 10 TeV can path through the dark star phase,
although the duration of this phase are very short. 
The final stellar masses are small for the dark stars with most massive WIMPs.

Note that we have used no-depletion models here for simplicity, because dark matter depletion
effect is small for the early stages of the dark star evolution.
If we take the DM depletion into account, a dark star ends the stable phase
earlier and above two quantities $M_{3.8}$ and $M_{4.0}$ become smaller.

\section{Conclusion}

We have studied the pre-main-sequence evolution of dark stars
by following the self-consistent stellar evolution.
We have suitably modified the code to incorporate 
the energy generation from spherically distributed DM.

Our base model with 
$m_{\chi}= 100 \ \mathrm{GeV}$ and $dM/dT=1.0 \times $
G-rate$ \ \mathrm{M_{\odot} \ yrs^{-1}}$ 
shows the characteristic features of the dark star phase; 
the large DM annihilation energy expands the star,
making the star to be in gravitational equilibrium.
The lower temperature is one of the peculiar properties
of the dark star.
This stable phase continues until the energy supply
from DM annihilation becomes insufficient to maintain the stable structure.
Finally, the star collapses rapidly and reaches the main-sequence phase.
At this point, the stellar mass has grown up 
to $M \sim 900 - 1000 \ \mathrm{M_{\odot}}$. 
Such features in dark star phase are all consistent with the findings of
previous works of \citet{iocco08, spolyar09}
who employed much simpler stellar models.

The dark star model has effectively two parameters.  
One of them is the gas mass accretion rate 
which determines the evolution of the gas and DM distribution. 
Cosmological simulations predict a variety of gas accretion rates.
For a small accretion rate, the period of 
the dark star phase increases 
whereas the final stellar mass decreases $M \sim 500 \ \mathrm{M_{\odot}}$.
Note however that
the mass is still larger than the standard Pop III (no-DM) case
($100 - 200\ \mathrm{M_{\odot}})$.
Another parameter is the DM particle mass 
which determines the energy generation rate. 
For a small DM particle mass,
the period of dark star phase becomes longer and the final stellar mass becomes larger
(see Table \ref{table5}). 

All models pass through the dark star phase and this phase is maintained by 
a little DM fuel which is less than $0.1 \ \%$ of the stellar mass. 
If the first stars in the universe have undergone such phase, 
there are exotic stars which are cool and massive in the early universe. 

Formation of very massive dark stars has an important implication.
Such stars eventually collapse gravitationally, to form massive
black holes with masses as large as $1000 \ M_{\odot}$.
The remnant black holes can also grow by accretion or by
mergers to seed super-massive black holes. 
The exotic feature of dark star's appearance, luminous and cool, 
may be able to use as the powerful clue to search such stars.
Our calculations show that first stars can grow to be dark stars with
luminosity $\sim \mathrm{few} \ \ 10^7 \ \mathrm{L_{\odot}}$ 
at most and they will not be detectable by {\sl JWST}.
However, \citet{freese10} estimate for {\sl super-massive dark stars (SMDSs)}
and such very massive and bright stars (about $10^{9-11} \ L_{\odot}$) can be detected by JWST.
\citet{sandick10} argue that
dark star remnants might survive to the present day 
in the Milky Way, leaving $\gamma$-ray signatures from DM annihilation. 

In future work, we will include the nuclear reaction to 
calculate the first star evolution from the dark star phase 
to the main-sequence phase completely. 
It would be also interesting to include capture of DM particles. 
Once the stars have contracted and stop growing in mass,
there will be no more 
DM supplied by adiabatic contraction.
However, DM particle may still be captured by the star.
Previous studies showed that this process can make the star back 
in the dark star phase again. 
We will explore the effect of DM capture
and make a complete evolutionary model of the first stars with DM annihilation.



\acknowledgments

This work was supported by the Grants-in-Aid for Scientific Research
(20041005, 20105004) from the MEXT of Japan 
and
the Grants-in-Aid for Young Scientists (S) 
20674003 by the Japan Society for the Promotion of Science, 
and support from World Premier International Research Center Initiative 
(WPI Initiative), MEXT, Japan.

\onecolumn

\clearpage

\begin{deluxetable}{rrrrrrr}
\tablewidth{0pt}
\tablecaption{Stellar properties for the ``base model'' calculation}
\tablehead{\colhead{$M_{\ast} \ [\mathrm{M_{\odot}}]$} & \colhead{$R_{\ast} \ [\mathrm{R_{\odot}}]$} & 
	\colhead{$L_{\ast} \ [\mathrm{L_{\odot}}]$} & \colhead{$T_{\mathrm{eff}} \ [\mathrm{K}]$} & 
	\colhead{$T_{\mathrm{cen}} \ [\mathrm{K}]$} & \colhead{$M_{\mathrm{DM}} \ [\mathrm{M_{\odot}}]$} & 
	\colhead{$t \ [\mathrm{Myr}]$}}
\startdata
200 & 1.51E3 & 1.83E6 & 5.49E3 & 4.69E5 & 0.166 & 0.017 \\
600 & 1.51E3 & 6.98E6 & 7.67E3 & 9.33E5 & 0.313 & 0.097 \\
800 & 1.19E3 & 1.08E7 & 9.65E3 & 1.42E6 & 0.287 & 0.153 \\
1000 & 21.98 & 3.24E7 & 9.33E4 & 9.25E7 & 3.75E-4 & 0.218 \\ 
\enddata
\tablecomments{
We tabulate the main results for the base dark star model at stellar masses, 
$M=200, \ 600, \ 800 \ \mathrm{and} \ 1000 \ \mathrm{M_{\odot}}$. 
The columns show, from left to right, the stellar radius, luminosity, surface 
temperature, central temperature, 
DM mass inside the star, and the time elapsed {\sl from the beginning of the calculation.}}
\label{table1}
\end{deluxetable}


\begin{deluxetable}{rrrrrrrr}
\tablewidth{0pt}
\tablecaption{Stellar properties for three accretion models at $T_{\mathrm{eff}}=10^4 \ \mathrm{K}$}
\tablehead{\colhead{Accretion Rate} & \colhead{$M_{\ast} \ [\mathrm{M_{\odot}}]$} & 
	\colhead{$R_{\ast} \ [\mathrm{R_{\odot}}]$} & \colhead{$L_{\ast} \ [\mathrm{L_{\odot}}]$} & 
	\colhead{$T_{\mathrm{eff}} \ [\mathrm{K}]$} & \colhead{$T_{\mathrm{cen}} \ [\mathrm{K}]$} & 
	\colhead{$M_{\mathrm{DM}} \ [\mathrm{M_{\odot}}]$} & \colhead{$t \ [\mathrm{Myr}]$}}
\startdata
1.0 $\times$ G-rate & 821 & 1.14E3 & 1.15E7 & 1.00E4 & 1.51E6 & 0.277 & 0.160 \\ 
0.5 $\times$ G-rate & 682 & 9.87E2 & 8.67E6 & 1.00E4 & 1.56E6 & 0.205 & 0.237 \\
0.2 $\times$ G-rate & 492 & 7.87E2 & 5.55E6 & 1.00E4 & 1.63E6 & 0.125 & 0.351 \\ 
\enddata
\tablecomments{
Stellar properties for three accretion models 
when the stellar surface temperature reaches $T_{\mathrm{eff}}=10^4 \ \mathrm{K}$. 
After this phase, the star begins to run out the DM fuel and begins to gravitationally contract.}
\label{table2}
\end{deluxetable}


\begin{deluxetable}{rrrrrrrr}
\tablewidth{0pt}
\tablecaption{Stellar properties for three accretion models at $T_{\mathrm{eff}}=10^5 \ \mathrm{K}$}
\tablehead{\colhead{Accretion Rate} & \colhead{$M_{\ast} \ [\mathrm{M_{\odot}}]$} & 
	\colhead{$R_{\ast} \ [\mathrm{R_{\odot}}]$} & \colhead{$L_{\ast} \ [\mathrm{L_{\odot}}]$} & 
	\colhead{$T_{\mathrm{eff}} \ [\mathrm{K}]$} & \colhead{$T_{\mathrm{cen}} \ [\mathrm{K}]$} & 
	\colhead{$M_{\mathrm{DM}} \ [\mathrm{M_{\odot}}]$} & \colhead{$t \ [\mathrm{Myr}]$}}
\startdata
1.0 $\times$ G-rate & 1002 & 19.27 & 3.32E7 & 1.00E5 & 1.07E8 & 2.43E-4 & 0.220 \\ 
0.5 $\times$ G-rate & 775 & 16.55 & 2.43E7 & 1.00E5 & 1.08E8 & 1.46E-4 & 0.291 \\
0.2 $\times$ G-rate & 534 & 13.44 & 1.62E7 & 1.00E5 & 1.09E8 & 7.65E-5 & 0.401 \\ 
\enddata
\tablecomments{
Stellar properties for three accretion models when the stellar surface temperature reaches 
$T_{\mathrm{eff}}=10^5 \ \mathrm{K}$. 
By this phase, the star has contracted sufficiently and the density and temperature has risen up 
enough to start the hydrogen burning. 
In this point, the star is not supported by DM annihilation energy, so we stop the calculation.}
\label{table3}
\end{deluxetable}


\begin{deluxetable}{rrrrrrrr}
\tablewidth{0pt}
\tablecaption{Stellar properties for some DM mass models at $T_{\mathrm{eff}}=10^4 \ \mathrm{K}$}
\tablehead{\colhead{$m_{\chi} \ [\mathrm{GeV}]$} & \colhead{$M_{\ast} \ [\mathrm{M_{\odot}}]$} & 
	\colhead{$R_{\ast} \ [\mathrm{R_{\odot}}]$} & \colhead{$L_{\ast} \ [\mathrm{L_{\odot}}]$} & 
	\colhead{$T_{\mathrm{eff}} \ [\mathrm{K}]$} & \colhead{$T_{\mathrm{cen}} \ [\mathrm{K}]$} & 
	\colhead{$M_{\mathrm{DM}} \ [\mathrm{M_{\odot}}]$} & \colhead{$t \ [\mathrm{Myr}]$}}
\startdata
1 & 1349 & 1.82E3 & 2.92E7 & 1.00E4 & 1.11E7 & 0.096 & 0.354 \\
10 & 1325 & 1.59E3 & 2.23E7 & 1.00E4 & 1.33E6 & 0.254 & 0.344 \\ 
20 & 1214 & 1.51E3 & 2.02E7 & 1.00E4 & 1.42E6 & 0.297 & 0.298 \\ 
50 & 1009 & 1.32E3 & 1.55E7 & 1.00E4 & 1.46E6 & 0.307 & 0.222 \\ 
100 & 821 & 1.14E3 & 1.15E7 & 1.00E4 & 1.51E6 & 0.277 & 0.160 \\ 
200 & 536 & 8.76E2 & 6.81E6 & 1.00E4 & 1.47E6 & 0.198 & 0.081 \\ 
\enddata
\tablecomments{
Stellar properties for DM mass variation models 
when stellar surface temperature reaches $T_{\mathrm{eff}}=10^4 \ \mathrm{K}$; 
the end of the stable dark star phase and the start of transformation into main-sequence star.}
\label{table4}
\end{deluxetable}


\begin{deluxetable}{rrrrrrrr}
\tablewidth{0pt}
\tablecaption{Stellar properties for some DM mass models at $T_{\mathrm{eff}}=10^5 \ \mathrm{K}$}
\tablehead{\colhead{$m_{\chi} \ [\mathrm{GeV}]$} & \colhead{$M_{\ast} \ [\mathrm{M_{\odot}}]$} & 
	\colhead{$R_{\ast} \ [\mathrm{R_{\odot}}]$} & \colhead{$L_{\ast} \ [\mathrm{L_{\odot}}]$} & 
	\colhead{$T_{\mathrm{eff}} \ [\mathrm{K}]$} & \colhead{$T_{\mathrm{cen}} \ [\mathrm{K}]$} & 
	\colhead{$M_{\mathrm{DM}} \ [\mathrm{M_{\odot}}]$} & \colhead{$t \ [\mathrm{Myr}]$}}
\startdata
1 & 1370 & 23.86 & 5.03E7 & 1.00E5 & 2.49E8 & 3.47E-6 & 0.362 \\
10 & 1381 & 23.32 & 4.83E7 & 1.00E5 & 1.07E8 & 4.96E-5 & 0.367 \\
20 & 1295 & 22.45 & 4.48E7 & 1.00E5 & 1.07E8 & 8.92E-5 & 0.333 \\
50 & 1138 & 20.85 & 3.86E7 & 1.00E5 & 1.07E8 & 1.59E-4 & 0.269 \\
100 & 1002 & 19.27 & 3.32E7 & 1.00E5 & 1.07E8 & 2.43E-4 & 0.220 \\
200 & 776 & 18.91 & 3.16E7 & 1.00E5 & 9.95E7 & 3.88E-4 & 0.146 \\
\enddata
\tablecomments{
Stellar properties for DM mass variation models 
when stellar surface temperature reaches $T_{\mathrm{eff}}=10^5 \ \mathrm{K}$; 
the end of calculation.}
\label{table5}
\end{deluxetable}




\end{document}